\documentclass[seceq]{ptptex}
\usepackage{amsmath,amscd}

\newcommand{\bbox}[1]{\boldsymbol #1}

\newcommand{\Tr}[1]{\mathrm{Tr} #1}
\newcommand{\Det}[1]{\mathrm{Det} #1}
\title{%
Euclidean Path Integral of the Gauge Field
}

\subtitle{Holomorphic Representation}    

\author{%
Seiji {\scshape Sakoda}%
}

\inst{%
Department of Applied Physics,
National Defence Academy, Hashirimizu\\
Yokosuka, 239-8686, Japan
}

\abst{Basing on the canonical quantization of a BRS invariant
Lagrangian, we construct holomorphic representation of path integrals
for Faddeev-Popov(FP) ghosts as well as for unphysical degrees of the
gauge field from covariant operator formalism. A thorough investigation
of a simple soluble gauge model with finite degrees will explain the
metric structure of the Fock space and constructions of path integrals
for quantized gauge fields with FP ghosts. We define fermionic coherent
states even for a Fock space equipped with indefinite metric to obtain
path integral representations of a generating functional and an
effective action. 
The same technique will also be developed for path integrals of
unphysical degrees in the gauge field to find complete correspondence,
that insures cancellation of FP determinant, between FP ghosts and
unphysical components of the gauge field. 
As a byproduct, we obtain an explicit form of Kugo-Ojima projection,
$P^{(n)}$, to the subspace with $n$-unphysical particles in terms of
creation and annihilation operators for the abelian gauge theory.}
\begin{document}

\maketitle

\section{Introduction}
Path integral quantization of a system with local gauge invariance is
beautifully formulated by the Faddeev-Popov(FP) procedure.\cite{FP} \ 
The characteristic feature of systems with such symmetries is the
presence of redundant gauge variant degrees of freedom required to
maintain manifest gauge invariance and also other symmetries such as
space-time covariance. 
Abelian and non-abelian gauge theories will be typical examples of such
systems. Physical contents of such theories can be extracted immediately
if we choose some gauge fixing condition to be able to eliminate
unphysical degrees of freedom at the cost of manifest covariance.
The covariant perturbation theory in terms of Feynman path
integral for quantum gauge theories therefore involves this price to be
paid. The FP procedure and its reinterpretation by the BRS
invariant formulation of a path integral of such systems(BFV-BRS
approach)\cite{BFV,BRST} play the role for this purpose. We shall call a
path integral formulated in this way, including original one by FP, as a
``conventional'' path integral in this paper. An important remark on the
conventional path integral of gauge theories is that the FP determinant
appears in this formalism must be understood as the absolute value of
the determinant because otherwise, in non-perturbative region of the
field configuration, the positivity of a field dependent functional
determinant is not guaranteed.\cite{TKcom} \ Therefore a path integral
that involves the FP determinant without taking its absolute value may
yield a meaningless result.

However beautiful, there exist several open questions in the formalism of
quantization by means of such prescription. Among them typical issue
will be the Gribov ambiguities.\cite{Gribov}\tocite{MN78} \ If we need to
analyze gauge theory beyond perturbation, the vanishing or sign changes
of the FP determinant will arise to result in obstructions for
the formulation of the theory in terms of path integral.
A prescription to modify the original
FP path integral, concerning the locality of the gauge fixing
factor and the compensating factor in the gauge space, was suggested to
recast the Gribov complication to the breaking of one to one
correspondence between the gauge parameter and the variable of the
functional integration of the normalization factor for a path
integral;\cite{KF79} \ but essential solution seems to be still far away
from our current knowledge about formulating gauge theories. 

Some alternatives for FP path integral have been proposed by a few
authors.\cite{KS,AFIO,FFKS,proj,VMV} \ In Ref.~\citen{KS}, \ Kashiwa and
Sakamoto formulated path integrals from operator formalism on the
reduced phase space for physical degrees then performed $c$-number
gauge transformations under a path integral with the aid of identities
with respect to delta functions or Gaussian integrations to find
manifestly covariant path integrals.
On the other hand, in Ref.~\citen{AFIO}, \ Arisue, Fujiwara, Inoue and
Ogawa dealt with the resolution of unity for vector space equipped with
indefinite metric as a fundamental ingredient. Then a path integral
based on the canonical formalism in the Feynman gauge for
electromagnetic field was derived. There seems to have been a
disagreement between the results of Ref.~\citen{KS} \ and
Ref.~\citen{AFIO} \ concerning the domain of integration with respect to
the time component of the vector field.

In addition to these attempts, another approach for constructing path
integrals within canonical formalism for systems with gauge invariance
have been proposed by others.\cite{FFKS,proj,VMV} \ In
Ref.~\citen{FFKS}, the authors made use of projection to the physical
subspace, since naive application of Faddeev-Senjanovic(FS)
formula\cite{FS} had seemed to fail to yield a well-defined path
integral for some Hamiltonian systems with first class constraints on
compact phase spaces.
The others assert that the projection method provides a prescription to
formulate path integrals avoiding the Gribov problem since it does not
rely on any gauge fixing condition.\cite{proj,VMV} \ 
The feature of the projection method being free from gauge
fixing will also help us to sidestep on the use of path integral with
FP ghosts. Therefore it will be best to formulate path
integrals for systems with gauge symmetry in terms of the projection
method if it can maintain the space-time covariance entirely, because the
connection to the operator formalism is very clear in this method.
As far as the present author knows, however, there seems to be no
satisfactory prescription for dealing with situations we meet in
quantization of gauge theories with a covariant gauge conditions within
this formalism. Hence it will be difficult for the projection method to
keep covariance manifest.

Another issue to be asked on the FP path integral may be the precise
definition of the path integral over ghost fermions. The original
definition of the ghost fermions is given by a formal path integral
representation of a functional determinant. 
If we adopt covariant type gauge fixing, we will
find a second order Lagrangian for the path integral of ghost
fermions. As is well-known, however, from coherent state path integral,
that is formulated basing upon canonical formalism, for usual physical
fermions,\cite{YOTK,berezin} \ there does not seem to exist any 
room for such Lagrangian in the action of path integrals for
fermions. Furthermore, the boundary condition for such fermion path
integral is not clear from the formal definition as a determinant. From
very definition, it may be naively expected to integrate whole of fermion
variables including those at both initial and final boundaries of a
Feynman kernel, otherwise the fermion path integral becomes a kernel
that contains wave function factor for ghosts instead of yielding a
determinant. Even so it is still unclear if we should integrate a
Feynman kernel with periodic or anti-periodic boundary condition. From
the BRS invariance point of view, it may be natural to expect the
periodic boundary conditions on all fields.\cite{HK1,KF96} \  This
immediately conflicts with the anti-periodic one for the determinant
from path integrals of usual fermions.\cite{YOTK} \
If it requires any boundary conditions other than the familiar
anti-periodic one, the path integral of ghost fermions then differs
significantly from the usual trace formula. In this sense the issue
itself shall be interesting as a subject of path integral technique.
These problems were solved by Kashiwa\cite{TKfermi} by constructing a
path integral of FP ghosts in the field diagonal representation. He also
clarified the relation of Kugo-Ojima's quartet mechanism\cite{KO} and the
Gaussian identity in the path integral. There may remain, however, a
room for considering alternatives for the path integral given by
Ref.~\citen{TKfermi} in terms of another representation.

Our main purpose in this article is therefore to formulate a path
integral of ghost fermions as well as the corresponding one of
unphysical degrees of the gauge field in a unified manner without making
use of FP prescription. To achieve this we introduce coherent states for
these unphysical variables. We shall try such procedure working with a
simple gauge invariant model in the BRS invariant treatment. In the next
section we explain our model and its quantization on representation
spaces with indefinite metric. Canonical quantization and path integral
representation for ghost fermions in terms of coherent states will be
investigated in section~\ref{sec:brs} in detail. In
section~\ref{sec:pi4gauge} we generalize the method of coherent state to
include unphysical degrees of the gauge field. Generating functionals
and effective actions both for FP ghosts and unphysical degrees of the
gauge field will then be evaluated in section~\ref{sec:gfefa}. Our
construction of coherent states and their use in gauge model requires
complete analysis of the vector space on which we formulate the quantum
theory of the model. Performing such a detailed investigation brings us
a nice understanding of BRS 
quartet\cite{KO} and Kugo-Ojima projection. We will show an explicit
form of Kugo-Ojima projection expressed as an integration to yield a
projection to an eigenspace of a number operator. This will be given in
section \ref{sec:kugoojima}. 
Application to the quantization of a free gauge field will be then
shortly discussed in section \ref{sec:maxwell}. We will confirm there
the applicability and reliability of our method by observing that our
prescription reproduces zeroth order results for covariant perturbation
of the quantized gauge field. Considerations on the relation of the results
of Ref.~\citen{KS} and Ref.~\citen{AFIO} will be made in
section~\ref{sec:discussion} by comparing our prescription in this paper
with the standard ``Euclidean Technique''.

\section{Simple models for the quantized gauge field}
\label{sec:toymodel}
In quantization of a free gauge field, considerations on the following
two Lagrangian systems are quite useful. For a fixed 3-dimensional
vector $\bbox{k}\ne0$, we take a Lagrangian 
\begin{equation}
 L_{\mathrm{phys}}=\frac{1}{2}\dot{\bbox{A}}^{2}-
 \frac{1}{2}k^{2}\bbox{A}^{2},\quad
 k=\vert\bbox{k}\vert,\
 \bbox{k}\cdot\bbox{A}=0
\label{eq:lagph}
\end{equation}
as a model for physical variables and
\begin{equation}
 L_{\mathrm{unphys}}=\frac{k^{2}}{2}(\dot{A}-A_{0})^{2}
\label{eq:lagup}
\end{equation}
for unphysical degrees.
The former for physical variables is
understood as a model for quantization in the Coulomb gauge while the
latter represents gauge variant degrees. This can be seen by observing
that the Lagrangian Eq.~\eqref{eq:lagup} has a gauge
symmetry under a gauge transformation
\begin{equation}
 A\mapsto A+\theta,\quad
 A_{0}\mapsto A_{0}+\dot{\theta}
\end{equation}
while there exists no such symmetry in the Lagrangian Eq.~\eqref{eq:lagph}.
Note that, in the above Lagrangians, the zero-modes both in physical and
unphysical degrees are excluded from our consideration. This corresponds
to a regularization for infrared singularities in quantization of the
massless gauge field.

If we introduce two independent vectors
$\bbox{e}_{s}(\bbox{k})$($s=1,\,2$) such that
\begin{equation}
 \bbox{k}\cdot\bbox{e}_{s}(\bbox{k})=0,\
 \bbox{e}_{s}(\bbox{k})\cdot\bbox{e}_{s{'}}(\bbox{k})=\delta_{ss{'}},
\end{equation}
the physical variables are parametrized as
\begin{equation}
 \bbox{A}(t)=\sum_{s=1,\,2}q_{s}(t;\bbox{k})\bbox{e}_{s}(\bbox{k})
\end{equation}
by two real variables $q_{s}(t;\bbox{k})$($s=1,\,2$). Then the
Lagrangian becomes
\begin{equation}
 L_{\mathrm{phys}}=\sum_{s=1,\,2}\frac{1}{2}
 \{\dot{q}_{s}^{2}(t;\bbox{k})-k^{2}q_{s}^{2}(t;\bbox{k})\}
\end{equation}
to be found as a set of two symmetric harmonic oscillators. Hence the
quantizations of physical variables are straightforward. We thus omit
detailed description of this procedure and concentrate the quantization
of unphysical degrees by writing $L_{\mathrm{unphys}}$ as $L$ for brevity.

\section{The BRS formalism}
\label{sec:brs}
In this section we consider canonical quantization of variables appear
in the BRS Lagrangian obtained from Eq.~\eqref{eq:lagup} by adding terms
both for gauge fixing and ghost fermions.

\subsection{BRS invariant Lagrangian}
\label{ssec:toymodel}
The BRS transforms of the original variables in Eq.~\eqref{eq:lagup} are
given by
\begin{equation}
 \bbox{\delta}_{B}A(t)=c(t),\
 \bbox{\delta}_{B}A_{0}(t)=\dot{c}(t),
\end{equation}
or by
\begin{equation}
 \delta_{B}A(t)=c(t)\theta,\
 \delta_{B}A_{0}(t)=\dot{c}(t)\theta,
\end{equation}
with a Grassmann parameter $\theta$.
Defining the BRS transform of the ghost $c(t)$ to be
\begin{equation}
  \bbox{\delta}_{B}c(t)=0,
\end{equation}
we can see the nil-potency of the BRS transformation on these
variables. Next we introduce the anti-ghost $\bar{c}(t)$ to satisfy
\begin{equation}
  \bbox{\delta}_{B}\bar{c}(t)=-iB(t).
\end{equation}
Then $\bbox{\delta}_{B}B(t)=0$ ensures the nil-potency of the BRS
transformation on all variables.
Taking these into account, the gauge fixing part of the Lagrangian is
defined by
\begin{equation}
 L_{\mathrm{GF}+\mathrm{FP}}=i\bbox{\delta}_{B}\left\{\bar{c}
 \left(\dot{A}_{0}+k^{2}A+\frac{\alpha}{2}B\right)\right\},
\end{equation}
for a \emph{covariant gauge condition}
\begin{equation}
 \dot{A}_{0}+k^{2}A+\alpha B=0.
\end{equation}
The reason why we call this as a covariant condition will be made clear
in the application of the present model to the gauge field in later
sections.

Explicitly, $L_{\mathrm{GF}+\mathrm{FP}}$ is given by
\begin{equation}
 L_{\mathrm{GF}+\mathrm{FP}}=B(\dot{A}_{0}+k^{2}A)+\frac{\alpha}{2}B^{2}+
 i\bar{c}(\ddot{c}+k^{2}c)
\end{equation}
to yield a gauge fixed Lagrangian
\begin{equation}
 \tilde{L}=L+L_{\mathrm{GF}+\mathrm{FP}}=
 \frac{k^{2}}{2}(\dot{A}-A_{0})^{2}+
 B(\dot{A}_{0}+k^{2}A)+\frac{\alpha}{2}B^{2}+
 i\bar{c}(\ddot{c}+k^{2}c).
\label{eq:lagbrs}
\end{equation}
Abelian nature of the symmetry under consideration allows us to separate
this total Lagrangian into two parts, the Lagrangian $L_{\mathrm{G}}$ that
describes unphysical degrees of the gauge field with the multiplier field
$B$, given by
\begin{equation}
 L_{\mathrm{G}}=\frac{k^{2}}{2}(\dot{A}-A_{0})^{2}+B(\dot{A}_{0}+k^{2}A)+
 \frac{\alpha}{2}B^{2},
 \label{eq:lab0}
\end{equation}
and the one $L_{\mathrm{FP}}$ for ghost fermions:
\begin{equation}
 L_{\mathrm{FP}}=i\bar{c}(\ddot{c}+k^{2}c).
 \label{eq:lfp0}
\end{equation}
The BRS transform of each Lagrangian is given by
\begin{equation}
 \bbox{\delta}_{B}L_{\mathrm{G}}=B(\ddot{c}+k^{2}c)
\end{equation}
and
\begin{equation}
 \bbox{\delta}_{B}L_{\mathrm{FP}}=-B(\ddot{c}+k^{2}c)
\end{equation}
respectively to result in the BRS invariance of the whole Lagrangian
$\tilde{L}$ by their cancellation.

\subsection{Coherent state path integral of ghost fermions}
\label{ssec:ghosts}
If we take care the BRS invariance of the total system, integration by
part on the kinetic term of the Lagrangian $L_{\mathrm{FP}}$ for FP
ghosts(See section \ref{sec:pi4gauge} for the consequence of this on the
corresponding part in the Lagrangian for the gauge field) can be done to
rewrite $L_{\mathrm{FP}}$ as
\begin{equation}
 L_{\mathrm{FP}}=-i\dot{\bar{c}}\dot{c}+
 ik^{2}\bar{c}c.
\end{equation}
The corresponding Hamiltonian is then given by
\begin{equation}
 H_{\mathrm{FP}}=ip_{\bar{c}}p_{c}-ik^{2}\bar{c}c
\label{eq:fpham0}
\end{equation}
in which $p_{c}$($p_{\bar{c}}$) being the canonical conjugate of
$c$($\bar{c}$). The canonical structure for the system in
consideration is defined by the following Poisson brackets:
\begin{equation}
 \{c,p_{c}\}=\{\bar{c},p_{\bar{c}}\}=1,\quad
 \{c,p_{\bar{c}}\}=\{\bar{c},p_{c}\}=
 \{c,\bar{c}\}=\{p_{c},p_{\bar{c}}\}=0.
\label{eq:pbra0}
\end{equation}
From equations of motion, we can set $c(t)$, $\bar{c}(t)$, $p_{c}(t)$
and $p_{\bar{c}}(t)$ as
\begin{equation}
\begin{gathered}
 c(t)=\frac{1}{\sqrt{2k}}(e^{-ikt}b+e^{ikt}b^{*}),\
 \bar{c}(t)=\frac{1}{\sqrt{2k}}(e^{-ikt}d+e^{ikt}d^{*}),\\
 p_{c}(t)=-\sqrt{\frac{k}{2}}(e^{-ikt}d-e^{ikt}d^{*}),\
 p_{\bar{c}}(t)=\sqrt{\frac{k}{2}}(e^{-ikt}b-e^{ikt}b^{*}).
\end{gathered}
\label{eq:mode0}
\end{equation}
Combining Eq.~\eqref{eq:pbra0} and Eq.~\eqref{eq:mode0} together, we
obtain Poisson brackets among $b$, $b^{*}$, $d$, and $d^{*}$:
\begin{equation}
\begin{gathered}
 \{b(t),d^{*}(t)\}=1,\ \{b^{*}(t),d(t)\}=-1,\ 
 \{b(t),b^{*}(t)\}=\{d(t),d^{*}(t)\}=0,\\
 \{b(t),b(t)\}=\{b^{*}(t),b^{*}(t)\}=
 \{d(t),d(t)\}=\{d(t),d^{*}(t)\}=0,
\end{gathered}
\end{equation}
in which $b(t)=e^{-ikt}b$, $b^{*}(t)=e^{ikt}b^{*}$, $d(t)=e^{-ikt}d$ and
$d^{*}(t)=e^{ikt}d^{*}$.

Upon quantization we replace these Poisson brackets by the following
anticommutators of Schr\"{o}dinger operators:
\begin{equation}
 \{\hat{b},\hat{d}^{\dagger}\}=i,\
 \{\hat{d},\hat{b}^{\dagger}\}=-i,\ 
 \text{ others }=0
\label{eq:acr0}
\end{equation}
or the equal-time anticommutators for Heisenberg operators
\begin{equation}
 \{\hat{b}(t),\hat{d}^{\dagger}(t)\}=i,\
 \{\hat{d}(t),\hat{b}^{\dagger}(t)\}=-i,\ 
 \text{ others }=0.
\label{eq:acr0h}
\end{equation}

To construct a representation of this algebra, let us introduce two
sorts of number operator, $\hat{N}_{i}(i=1,\,2)$, by
\begin{equation}
 \hat{N}_{1}=i\hat{b}^{\dagger}\hat{d},\
 \hat{N}_{2}=-i\hat{d}^{\dagger}\hat{b}
\end{equation}
to find
\begin{gather}
 [\hat{N}_{1},\hat{b}^{\dagger}]=\hat{b}^{\dagger},\
 [\hat{N}_{1},\hat{d}]=-\hat{d},\
 [\hat{N}_{1},\hat{b}]=[\hat{N}_{1},\hat{d}^{\dagger}]=0,\\
 [\hat{N}_{2},\hat{b}]=-\hat{b},\
 [\hat{N}_{2},\hat{d}^{\dagger}]=\hat{d}^{\dagger},\
 [\hat{N}_{2},\hat{b}^{\dagger}]=[\hat{N}_{2},\hat{d}]=0.
\end{gather}
Apparently $\{\hat{N}_{1},\,\hat{b}^{\dagger},\,\hat{d}\}$ and
$\{\hat{N}_{2},\,\hat{d}^{\dagger},\,\hat{b}\}$ form their own closed
algebra. Hence we obtain the following basis vectors of the
representation space:
\begin{equation}
 \hat{N}_{i}\vert\{n_{2}n_{1}\}\rangle=n_{i}\vert\{n_{2}n_{1}\}\rangle,\
 n_{i}=0,\,1\ (i=1,\,2)
\end{equation}
where the order of $n_{i}$ in $\vert\{n_{2}n_{1}\}\rangle$ has meaning
and the vacuum $\vert\{00\}\rangle$ is defined by
\begin{equation}
 \hat{b}\vert\{00\}\rangle=\hat{d}\vert\{00\}\rangle=0,
\end{equation}
upon which the rest of the basis are generated as
\begin{equation}
 \vert\{01\}\rangle=\hat{b}^{\dagger}\vert\{00\}\rangle,\
 \vert\{10\}\rangle=\hat{d}^{\dagger}\vert\{00\}\rangle,\
 \vert\{11\}\rangle=\hat{d}^{\dagger}\hat{b}^{\dagger}\vert\{00\}\rangle.
\end{equation}
We may naturally expect the vacuum $\vert\{00\}\rangle$ to have positive
norm and to be normalized as
\begin{equation}
 \langle\{00\}\vert\{00\}\rangle=1.
\end{equation}
Then by writing the hermitian conjugation of a state vector without
taking the metric into account as
\begin{equation}
 \langle\{n_{2}n_{1}\}\vert=(\vert\{n_{2}n_{1}\}\rangle)^{\dagger},
\end{equation}
we find the {\em norm} of others given by
\begin{equation}
 \langle\{01\}\vert\{01\}\rangle=
 \langle\{10\}\vert\{10\}\rangle=0,\
 \langle\{11\}\vert\{11\}\rangle=-1.
\end{equation}
Though one particle states, $\vert\{01\}\rangle$ and
$\vert\{10\}\rangle$ are zero-normed, they have non-zero inner products
with each other
\begin{equation}
 \langle\{01\}\vert\{10\}\rangle=i \text{ and }
 \langle\{10\}\vert\{01\}\rangle=-i
\end{equation}
to yield off diagonal elements to the metric of this vector space.
Taking this metric structure into account, we find the identity operator on this vector space given by
\begin{equation}
 1=\vert\{00\}\rangle\langle\{00\}\vert+
 i\vert\{01\}\rangle\langle\{10\}\vert-
 i\vert\{10\}\rangle\langle\{01\}\vert-
 \vert\{11\}\rangle\langle\{11\}\vert.
\label{eq:identity}
\end{equation}
If we introduce a conjugation defined by
\begin{equation}
 \vert\{n_{2}n_{1}\}\rangle\leftrightarrow
 \langle\underline{\{n_{2}n_{1}\}}\vert=
 \langle\{n_{2}n_{1}\}\vert\hat{\eta}_{\mathrm{FP}},\
 \hat{\eta}_{\mathrm{FP}}=
 e^{\pi
 i(\hat{b}^{\dagger}+i\hat{d}^{\dagger})(\hat{b}-i\hat{d})/2}\quad
 (\hat{\eta}^{-1}_{\mathrm{FP}}=
 \hat{\eta}^{\dagger}_{\mathrm{FP}}=
 \hat{\eta}_{\mathrm{FP}})
\end{equation}
corresponding to the above metric structure, the identity operator can
be expressed in a concise form by
\begin{equation}
 1=\sum_{n_{i}=0,1}\vert\{n_{2}n_{1}\}\rangle
 \langle\underline{\{n_{2}n_{1}\}}\vert.
\label{eq:identity0}
\end{equation}
Note here that the conjugation introduced above generates the following
transformation on operators:
\begin{equation}
 (\hat{b},\hat{d})\mapsto
 \hat{\eta}_{\mathrm{FP}}^{\dagger}(\hat{b},\hat{d})\hat{\eta}_{\mathrm{FP}}=
 (i\hat{d},-i\hat{b})
\end{equation}
together with their hermitian conjugations.

We regard the representation space to be a Grassmann valued vector space
in order to construct coherent states\cite{YOTK} of the above fermions.
Taking the algebra above into account, we define a coherent state
\begin{equation}
 \vert\bbox{\xi}\rangle=
 e^{i(\hat{b}^{\dagger}\xi_{2}-\hat{d}^{\dagger}\xi_{1})}\vert\{00\}\rangle=
 \vert\{00\}\rangle-i\xi_{2}\vert\{01\}\rangle+i\xi_{1}\vert\{10\}\rangle+
 \xi_{2}\xi_{1}\vert\{11\}\rangle
\end{equation}
to be a simultaneous eigenvector of $\hat{b}$ and $\hat{d}$:
\begin{equation}
 \hat{b}\vert\bbox{\xi}\rangle=\xi_{1}\vert\bbox{\xi}\rangle,\
 \hat{d}\vert\bbox{\xi}\rangle=\xi_{2}\vert\bbox{\xi}\rangle.
\end{equation}
Taking hermitian conjugation of the above equations, we find that
$\hat{b}^{\dagger}$ and $\hat{d}^{\dagger}$ have common left
eigenvectors defined by
\begin{equation}
 \langle\bbox{\xi}^{*}\vert=\langle\{00\}\vert 
 e^{-i(\xi_{2}^{*}\hat{b}-\xi_{1}^{*}\hat{d})}=
 \langle\{00\}\vert+i\langle\{01\}\vert\xi_{2}^{*}-
 i\langle\{10\}\vert\xi_{1}^{*}+
 \langle\{11\}\vert\xi_{1}^{*}\xi_{2}^{*}
\end{equation}
to yield
\begin{equation}
 \langle\bbox{\xi}^{*}\vert\bbox{\xi}^{\prime}\rangle=
 e^{i(\xi_{1}^{*}\xi_{2}^{\prime}-\xi_{2}^{*}\xi_{1}^{\prime})}.
\label{eq:fpinpro0}
\end{equation}
The set of coherent states thus defined is complete:
\begin{equation}
 \int\!d^{2}\xi^{*}\,d^{2}\xi\,
 e^{-i(\xi_{1}^{*}\xi_{2}-\xi_{2}^{*}\xi_{1})}
 \vert\bbox{\xi}\rangle\langle\bbox{\xi}^{*}\vert=1,
\label{eq:coh0}
\end{equation}
where use has been made of our convention of integration with respect to
Grassmann numbers\cite{YOTK}
\begin{equation}
 \int\!\xi\,d\xi=\int\!\xi^{*}\,d\xi^{*}=i,\quad
 \int\!d\xi=\int\!d\xi^{*}=0.
\end{equation}
The order of variables in the measure will be important for integrations
over Grassmann numbers; Explicit definition of the measure in
Eq.~\eqref{eq:coh0} is given  by
\begin{equation}
 d^{2}\xi^{*}=d\xi^{*}_{1}\,d\xi^{*}_{2},\
 d^{2}\xi=d\xi_{1}\,d\xi_{2}.
\end{equation}

If we make use of another version of the dual vector to
$\vert\bbox{\xi}\rangle$, given by
\begin{equation}
 \langle\underline{\bbox{\xi}}\vert=
 \langle\{00\}\vert e^{\xi_{1}^{*}\hat{b}+\xi_{2}^{*}\hat{d}}=
 \langle\bbox{\xi}^{*}\vert\hat{\eta}_{\mathrm{FP}},
\end{equation}
the inner product Eq.~\eqref{eq:fpinpro0} and the expression for
the resolution of unity Eq.~\eqref{eq:coh0} given above are replaced by
\begin{equation}
 \langle\underline{\bbox{\xi}}\vert\bbox{\xi}^{\prime}\rangle=
 e^{\bbox{\xi}^{\dagger}\bbox{\xi}{'}}
\label{eq:fpinpro1}
\end{equation}
and to
\begin{equation}
 \int\!(d\xi\,d\xi^{*})^{2}\,
 e^{-\bbox{\xi}^{\dagger}\bbox{\xi}}
 \vert\bbox{\xi}\rangle\langle\underline{\bbox{\xi}}\vert=1,\
 (d\xi\,d\xi^{*})^{2}=
 d\xi_{1}\,d\xi^{*}_{1}\,d\xi_{2}\,d\xi^{*}_{2}
\label{eq:coh1}
\end{equation}
respectively.
Though above two expressions of the resolution of unity are equivalent,
it will be natural to use the latter because it reflects the metric
structure and yields a Gaussian path integral below. Furthermore, it is
this latter form that has an analogue in the quantization of unphysical
degrees of gauge fields. We thus utilize the resolution of unity
Eq.~\eqref{eq:coh1} in the following to obtain a holomorphic representation
of a path integral for ghost fermions.

Now we proceed to formulate a path integral for the system in terms of
coherent states defined above. The Hamiltonian operator
\begin{equation}
 \hat{H}_{\mathrm{FP}}=ik(\hat{b}^{\dagger}\hat{d}-\hat{d}^{\dagger}\hat{b})-k
 =k(\hat{N}_{\mathrm{FP}}-1),\
 \hat{N}_{\mathrm{FP}}=\hat{N}_{1}+\hat{N}_{2}
\label{eq:fpham1}
\end{equation}
for the quantum system is obtained by replacing canonical variables with
corresponding operators: 
\begin{equation}
 \hat{c}=\frac{1}{\sqrt{2k}}(\hat{b}+\hat{b}^{\dagger}),\
 \hat{\bar{c}}=\frac{1}{\sqrt{2k}}(\hat{d}+\hat{d}^{\dagger}),\
 \hat{p}_{c}=-\sqrt{\frac{k}{2}}(\hat{d}-\hat{d}^{\dagger}),\
 \hat{p}_{\bar{c}}=\sqrt{\frac{k}{2}}(\hat{b}-\hat{b}^{\dagger})
\end{equation}
in the classical one given by Eq.~\eqref{eq:fpham0}.
A Feynman kernel for an imaginary time $\beta$ is defined by
\begin{equation}
 \langle\underline{\bbox{\xi}_{F}}\vert
 e^{-\beta\hat{H}_{\mathrm{FP}}}\vert\bbox{\xi}_{I}\rangle=
 \lim\limits_{n\to\infty}\langle\underline{\bbox{\xi}_{F}}\vert
 \left(1-\epsilon\hat{H}_{\mathrm{FP}}\right)^{n}
 \vert\bbox{\xi}_{I}\rangle,\quad
 \epsilon=\frac{\beta}{n}.
\end{equation}
Repeated use of the resolution of unity given by Eq.~\eqref{eq:coh1} and
the evaluation of the matrix element
\begin{equation}
 \langle\underline{\bbox{\xi}(j)}\vert
 \left(1-\epsilon\hat{H}_{\mathrm{FP}}\right)
 \vert\bbox{\xi}(j-1)\rangle=
 \exp\left\{{(1-\epsilon k)\bbox{\xi}^{\dagger}(j)\bbox{\xi}(j-1)
 +\epsilon k}\right\}
\end{equation}
will bring us a path integral formula
\begin{multline}
 K(\bbox{\xi}_{F},\bbox{\xi}_{I};\beta)=e^{\beta k}\,
 \lim\limits_{n\to\infty}\int\prod_{i=1}^{n-1}
 (d\xi(i)\,d\xi^{*}(i))^{2}\\
 \times \exp\left\{{-\sum_{j=1}^{n-1}
 \bbox{\xi}^{\dagger}(j)\bbox{\xi}(j)+(1-\epsilon k)\sum_{j=1}^{n}
 \bbox{\xi}^{\dagger}(j)\bbox{\xi}(j-1)}\right\}
\end{multline}
for the Euclidean kernel 
\begin{equation}
 K(\bbox{\xi}_{F},\bbox{\xi}_{I};\beta)=
 \langle\underline{\bbox{\xi}_{F}}\vert
 e^{-\beta\hat{H}_{\mathrm{FP}}}\vert\bbox{\xi}_{I}\rangle.
\end{equation}
Note that it is the expression in Eq.~\eqref{eq:coh1} for the resolution of
unity that brings a Gaussian path integral above even for the present
system equipped with indefinite metric.

Gaussian integrations are carried out by use of the standard technique
for Grassmann variables to yield
\begin{equation}
\begin{aligned}
 K(\bbox{\xi}_{F},\bbox{\xi}_{I};\beta)=&
 e^{\beta k}\lim\limits_{n\to\infty}
 \exp\left\{{(1-\epsilon k)^{n}
 \bbox{\xi}_{F}^{\dagger}\bbox{\xi}_{I}}\right\}\\
 =&
 \exp\left({\beta k+e^{-\beta k}
 \bbox{\xi}_{F}^{\dagger}\bbox{\xi}_{I}}\right).
\label{eq:kern1}
\end{aligned}
\end{equation}
The validity of our construction of Euclidean path integral for
FP ghosts can be checked by an observation
\begin{equation}
 \lim\limits_{\beta\to\infty}e^{-\beta k}
 K(\bbox{\xi}_{F},\bbox{\xi}_{I};\beta)=1=
 \langle\underline{\bbox{\xi}_{F}}\vert\{00\}\rangle
 \langle\{00\}\vert\bbox{\xi}_{I}\rangle.
\end{equation}
We thus formulated a holomorphic representation of path integral for
ghost fermions.

\section{Path integral of unphysical degrees of the gauge field}
\label{sec:pi4gauge}
Our task in this section is to perform path integral of the Lagrangian
\begin{equation}
 L_{\mathrm{G}}=\frac{k^{2}}{2}(\dot{A}-A_{0})^{2}
 -\dot{B}A_{0}+k^{2}AB+\frac{\alpha}{2}B
\end{equation}
for unphysical components, $A$ and $A_{0}$, of the gauge field and the
multiplier field $B$. Here, a comment is in order. In the above
Lagrangian, we have performed an integration by parts for the term
$B\dot{A}_{0}$ in the original Lagrangian Eq.~\eqref{eq:lab0}. It is
necessary to preserve the BRS invariance of the Lagrangian because we
have already done the same on the term $i\bar{c}\ddot{c}$ in the FP
ghost part. Changing $i\bar{c}\ddot{c}$ to $-i\dot{\bar{c}}\dot{c}$
requires addition of $d(-i\bar{c}\dot{c})/dt$ which is a half of the BRS
exact total derivative $d\{\bbox{\delta}_{B}(-i\bar{c}A_{0})\}/dt$. We
must therefore add this total derivative to the total Lagrangian in
order to maintain the BRS invariance.\cite{KO}\

\subsection{Canonical formulation for covariant gauge conditions}
\label{ssec:ccr4gauge}
Canonical momenta of the gauge and multiplier fields are given by
\begin{equation}
 P=k^{2}(\dot{A}-A_{0}),\
 P_{0}=0,\
 P_{B}=-A_{0}
\end{equation}
and the Hamiltonian for this system reads
\begin{equation}
 H_{\mathrm{G}}=\frac{1}{2k^{2}}P^{2}+A_{0}P-k^{2}AB
 -\frac{\alpha}{2}B^{2},
\label{eq:gham}
\end{equation}
where $A$ and $B$ are regarded as coordinate variables with their
canonical conjugates $P=-\dot{B}$ and $P_{B}=-A_{0}$ as a consequence of the
Dirac-Bergmann prescription.\cite{dirac} \
By use of equations of motion, $P$ can be
always replaced by $-\dot{B}$. 
Hence we have fundamental commutation relations
\begin{equation}
 [\hat{A}(t),-\dot{\hat{B}}(t)]=i,\
 [\hat{B}(t),-\hat{A}_{0}(t)]=i
\label{eq:eqcm0}
\end{equation}
as well as
\begin{equation}
 [\hat{A}(t),\hat{A}_{0}(t)]=[\hat{A}(t),\hat{B}(t)]=
 [\hat{A}_{0}(t),\dot{\hat{B}}(t)]=[\hat{B}(t),\dot{\hat{B}}(t)]=0
\label{eq:eqcm1}
\end{equation}
together with the Heisenberg equations
\begin{align}
\label{eq:eqofm0}
 \ddot{\hat{A}}(t)-\dot{\hat{A}}_{0}(t)-\hat{B}(t)&=0,\\
\label{eq:eqofm1}
 k^{2}(\dot{\hat{A}}(t)-\hat{A}_{0}(t))+\dot{\hat{B}}(t)&=0,\\
\label{eq:eqofm2}
 \dot{\hat{A}}_{0}+k^{2}\hat{A}(t)+\alpha \hat{B}(t)&=0,
\end{align}
for quantization. As can be seen easily, $\hat{B}(t)$ obeys a free field
equation:
\begin{equation}
 \ddot{\hat{B}}+k^{2}\hat{B}(t)=0
\end{equation}
while $\hat{A}(t)$ and $\hat{A}_{0}(t)$ in general involves dipole ghost
and satisfy
\begin{equation}
 \ddot{\hat{A}}(t)+k^{2}\hat{A}(t)=(1-\alpha)\hat{B}(t),\
 \ddot{\hat{A}}_{0}(t)+k^{2}\hat{A}_{0}(t)=(1-\alpha)\dot{\hat{B}}(t).
\end{equation}

In order to accomplish the quantization of the system,
we need to find a representation for the above algebra.
To this aim, we first study formal solutions of the Heisenberg equations
for $\hat{A}(t)$, $\hat{A}_{0}(t)$ and $\hat{B}(t)$ by making use of the
quantum mechanical version of the formulation for quantizing vector
fields with indefinite metric\cite{FK,NL,NN}. \ 
If we introduce Fourier transforms of Heisenberg operators by
\begin{align}
 \label{eq:ftab}
 \hat{A}(t)=&\int_{-\infty}^{\infty}\!\!dp\,
 \{\hat{a}(p)e^{-ipt}+\hat{a}^{\dagger}(p)e^{ipt}\},\\
 \hat{A}_{0}(t)=&\int_{-\infty}^{\infty}\!\!dp\,
 \{\hat{a}_{0}(p)e^{-ipt}+\hat{a}_{0}^{\dagger}(p)e^{ipt}\},\\
 \hat{B}(t)=&\int_{-\infty}^{\infty}\!\!dp\,
 \{\hat{\beta}(p)e^{-ipt}+\hat{\beta}^{\dagger}(p)e^{ipt}\},
\end{align}
equations of motion will be formally solved by putting
\begin{align}
 \label{eq:iftab}
 \hat{a}(p)=&-i\theta(p)\left\{\delta(p^{2}-k^{2})
 \| e^{ipt},\hat{A}(t)\|+
 (1-\alpha)\delta{'}(p^{2}-k^{2})\| e^{ipt},\hat{B}(t)\|\right\},\\
 \hat{a}_{0}(p)=&-i\theta(p)\left\{\delta(p^{2}-k^{2})
 \| e^{ipt},\hat{A}_{0}(t)\|+
 (1-\alpha)\delta{'}(p^{2}-k^{2})\| e^{ipt},\dot{\hat{B}}(t)\|\right\},\\
 \hat{\beta}(p)=&-i\theta(p)\delta(p^{2}-k^{2})\| e^{ipt},\hat{B}(t)\|
\end{align}
in which $\| F(t),G(t)\|$ being the Wronskian of $F(t)$ and $G(t)$.

As the most fundamental ingredient for the canonical formulation, we
require the existence of the vacuum state $\vert0\rangle$ to fulfill
\begin{equation}
 \hat{a}(p)\vert0\rangle=\hat{a}_{0}(p)\vert0\rangle=
 \hat{\beta}(p)\vert0\rangle=0
 \label{eq:vac0}
\end{equation}
or equivalently
\begin{equation}
 \hat{A}^{(+)}(t)\vert0\rangle=
 \hat{A}_{0}^{(+)}(t)\vert0\rangle=
 \hat{B}^{(+)}(t)\vert0\rangle=0,
 \label{eq:vac1}
\end{equation}
where positive and negative frequency parts of the Heisenberg operators
are defined by
\begin{equation}
 (\hat{A}^{(+)}(t),\hat{A}_{0}^{(+)}(t),\hat{B}^{(+)}(t))=
 \int_{-\infty}^{\infty}\!\!dp\,
 (\hat{a}(p),\hat{a}_{0}(p),\hat{\beta}(p))e^{-ipt},
\label{eq:pos}
\end{equation}
and by
\begin{equation}
 (\hat{A}^{(-)}(t),\hat{A}_{0}^{(-)}(t),\hat{B}^{(-)}(t))=
 \int_{-\infty}^{\infty}\!\!dp\,
 (\hat{a}^{\dagger}(p),\hat{a}_{0}^{\dagger}(p),
 \hat{\beta}^{\dagger}(p))e^{ipt},
\label{eq:neg}
\end{equation}
respectively.

By making use of the equal-time commutators in Eq.~\eqref{eq:eqcm0} and
Eq.~\eqref{eq:eqcm1}, we can find the following commutation relations for
the Fourier coefficients:
\begin{align}
 [\hat{a}(p),\hat{a}^{\dagger}(q)]=&\theta(p)\delta(p-q)
 \frac{1}{k^{2}}\left\{\delta(p^{2}-k^{2})-
 (1-\alpha)k^{2}\delta{'}(p^{2}-k^{2})\right\},\\
 [\hat{a}_{0}(p),\hat{a}_{0}^{\dagger}(q)]=&-\theta(p)\delta(p-q)
 \left\{\delta(p^{2}-k^{2})+
 (1-\alpha)k^{2}\delta{'}(p^{2}-k^{2})\right\},\\
 [\hat{\beta}(p),\hat{\beta}^{\dagger}(q)]=&0,\\
 [\hat{a}(p),\hat{b}^{\dagger}(q)]=&[\hat{\beta}(p),\hat{a}^{\dagger}(q)]=
 -\theta(p)\delta(p-q)\delta(p^{2}-k^{2}),\\
 [\hat{a}_{0}(p),\hat{\beta}^{\dagger}(q)]=&-[\hat{\beta}(p),\hat{a}_{0}^{\dagger}(q)]=
 ik\theta(p)\delta(p-q)\delta(p^{2}-k^{2}),\\
 [\hat{a}(p),\hat{a}_{0}^{\dagger}(q)]=&-[\hat{a}_{0}(p),\hat{a}^{\dagger}(q)]
 = -i(1-\alpha)\theta(p)\delta(p-q)p\delta{'}(p^{2}-k^{2}).
\label{eq:covcom}
\end{align}
These relations can be utilized for determination of arbitrary
commutators of Heisenberg operators and Green's functions.

\subsection{Coherent states for unphysical degrees of the gauge field}
\label{ssec:cs4gauge}
We here define coherent states for unphysical degrees of the gauge
field. To this aim, we rewrite Heisenberg operators in
Eq.~\eqref{eq:pos} as follows
\begin{equation}
 \hat{A}^{(+)}(t)=\frac{1}{\sqrt{2k}}\hat{\mathrm{a}}(t),\
 \hat{A}_{0}^{(+)}(t)=\frac{1}{\sqrt{2k}}\hat{\mathrm{a}}_{0}(t),\
 \hat{B}^{(+)}(t)=\frac{1}{\sqrt{2k}}\hat{\mathrm{b}}(t),
\end{equation}
and their hermitian conjugates in Eq.~\eqref{eq:neg} as well to find the
following equal-time commutation relations
\begin{equation}
\begin{gathered}
{} [\hat{\mathrm{a}}(t),\hat{\mathrm{a}}^{\dagger}(t)]=
 \frac{1+\alpha}{2k^{2}},\
 [\hat{\mathrm{a}}_{0}(t),\hat{\mathrm{a}}_{0}^{\dagger}(t)]=
 -\frac{1+\alpha}{2},\
 [\hat{\mathrm{a}}(t),\hat{\mathrm{a}}_{0}^{\dagger}(t)]=0,\\
 [\hat{\mathrm{a}}(t),\hat{\mathrm{b}}^{\dagger}(t)]=-1,\
 [\hat{\mathrm{a}}_{0}(t),\hat{\mathrm{b}}^{\dagger}(t)]=ik
\label{eq:gcom0}
\end{gathered}
\end{equation}
from Eq.~\eqref{eq:covcom}.
For constructing a representation of this algebra, it is convenient to
introduce the following Heisenberg operators:
\begin{equation}
\label{eq:galgop}
 \hat{\mathcal{B}}(t)=\frac{1}{k}\hat{\mathrm{b}}(t),\
 \hat{\mathcal{D}}(t)=\frac{i}{2}(k\hat{\mathrm{a}}(t)+
 i\hat{\mathrm{a}}_{0}(t)),\
 \hat{\bar{\mathcal{D}}}(t)=\frac{i}{2}(k\hat{\mathrm{a}}(t)-
 i\hat{\mathrm{a}}_{0}(t))
\end{equation}
with their hermitian conjugates. By use of the commutation relations in
Eq.~\eqref{eq:gcom0}, we obtain
\begin{equation}
\label{eq:galg0}
 [\hat{\mathcal{B}}(t),\hat{\mathcal{D}}^{\dagger}(t)]=i,\
 [\hat{\mathcal{D}}(t),\hat{\mathcal{B}}^{\dagger}(t)]=-i,
\end{equation}
as well as
\begin{equation}
\label{eq:galg1}
 [\hat{\mathcal{B}}(t),\hat{\mathcal{B}}^{\dagger}(t)]=
 [\hat{\mathcal{D}}(t),\hat{\mathcal{D}}^{\dagger}(t)]=
 [\hat{\mathcal{B}}(t),\hat{\mathcal{D}}(t)]= 0
\end{equation}
for $\hat{\mathcal{B}}(t)$ and $\hat{\mathcal{D}}(t)$ and their conjugates.
As for the operator $\hat{\bar{\mathcal{D}}}(t)$, we can utilize the relation:
\begin{equation}
 \hat{\bar{\mathcal{D}}}(t)=-\frac{i}{4}(1+\alpha)\hat{\mathcal{B}}(t)
\label{eq:soldbar}
\end{equation}
by observation that equal-time commutators of $\hat{\bar{\mathcal{D}}}(t)$ and
$\hat{\bar{\mathcal{D}}}^{\dagger}(t)$ with other operators vanish excepting
\begin{equation}
 [\hat{\mathcal{D}}(t),\hat{\bar{\mathcal{D}}}^{\dagger}(t)]=-
 [\hat{\bar{\mathcal{D}}}(t),\hat{\mathcal{D}}^{\dagger}(t)]=
 \frac{1+\alpha}{4}.\
\end{equation}
Note that the relation Eq.~\eqref{eq:soldbar} can be also obtained directly by
considering the Fourier transform of
\begin{equation}
 -ip\hat{a}_{0}(p)+k^{2}\hat{a}(p)=-\alpha\hat{\beta}(p)
\end{equation}
which is a consequence of the equations of motion(Eq.~\eqref{eq:eqofm2}).
Inversion of the definition of the operators $\hat{\mathcal{B}}(t)$ and
$\hat{\mathcal{D}}(t)$ can be done to express original Heisenberg operators in
terms of them if we also make a use of Eq.~\eqref{eq:soldbar}.
We thus obtain
\begin{equation}
 \begin{aligned}
 \hat{A}(t)=&-\frac{1}{\sqrt{2k^{3}}}\left\{
 \frac{1+\alpha}{4}(\hat{\mathcal{B}}(t)+\hat{\mathcal{B}}^{\dagger}(t))+
 i(\hat{\mathcal{D}}(t)-\hat{\mathcal{D}}^{\dagger}(t))\right\},\\
 \hat{A}_{0}(t)=&-\frac{i}{\sqrt{2k}}\left\{
 \frac{1+\alpha}{4}(\hat{\mathcal{B}}(t)-\hat{\mathcal{B}}^{\dagger}(t))-
 i(\hat{\mathcal{D}}(t)+\hat{\mathcal{D}}^{\dagger}(t))\right\},\\
 \hat{B}(t)=&\sqrt{\frac{k}{2}}(
 \hat{\mathcal{B}}(t)+\hat{\mathcal{B}}^{\dagger}(t)),\
 \dot{\hat{B}}(t)=-i\sqrt{\frac{k^{3}}{2}}(
 \hat{\mathcal{B}}(t)-\hat{\mathcal{B}}^{\dagger}(t)).
\label{eq:reverse}
 \end{aligned}
\end{equation}
As must be clear from definition, $\hat{\mathcal{B}}(t)$ is essentially
the positive frequency part of $\hat{B}(t)$ and hence BRS invariant. On the
other hand, though the form of the definition is independent of
$\alpha$, the operator $\hat{\mathcal{D}}(t)$ is a gauge dependent objects
through the $\alpha$-dependence of $\hat{\mathrm{a}}(t)$ and
$\hat{\mathrm{a}}_{0}(t)$.

Interestingly, commutation relations above closely resemble the
anticommutation relations Eq.~\eqref{eq:acr0} for ghost fermions. The
similarity is not limited only to the algebra but also seen in the form
of Hamiltonian for both systems. If we express the Hamiltonian
$\hat{H}_{\mathrm{G}}$ in terms of above operators by making use of
Eq.~\eqref{eq:reverse}, we find
\begin{equation}
\begin{aligned}
 \hat{H}_{\mathrm{G}}=&\hat{H}_{\mathrm{G}}{'}+
 \frac{1-\alpha}{2}k\hat{\mathcal{B}}^{\dagger}(t)\hat{\mathcal{B}}(t),\\
 \hat{H}_{\mathrm{G}}{'}=
 &ik(\hat{\mathcal{B}}^{\dagger}(t)\hat{\mathcal{D}}(t)-
 \hat{\mathcal{D}}^{\dagger}(t)\hat{\mathcal{B}}(t))+k.
\end{aligned}
\end{equation}
Combining this Hamiltonian together with that for ghost fermions, we see
that total Hamiltonian for the system under consideration is expressed
as
\begin{equation}
 \hat{H}=
 ik\{(\hat{\mathcal{B}}^{\dagger}(t)\hat{\mathcal{D}}(t)-
 \hat{\mathcal{D}}^{\dagger}(t)\hat{\mathcal{B}}(t))+
 (\hat{b}^{\dagger}(t)\hat{d}(t)-\hat{d}^{\dagger}(t)\hat{b}(t))\}+
 \frac{1-\alpha}{2}k\hat{\mathcal{B}}^{\dagger}(t)\hat{\mathcal{B}}(t),
\label{eq:totham}
\end{equation}
where $\hat{b}(t)$, $\hat{d}(t)$ and their hermitian conjugates designate Heisenberg
operators for ghost fermions. The last term, proportional to $1-\alpha$,
in the Hamiltonian can be eliminated by shifting $\hat{\mathcal{D}}(t)$ and
its conjugate according to
\begin{equation}
 \label{eq:dshift}
 \hat{\mathcal{D}}(t)\mapsto\hat{\mathcal{D}}{'}(t)=
 \hat{\mathcal{D}}(t)-\frac{i}{4}(1-\alpha)\hat{\mathcal{B}}(t),\
 \hat{\mathcal{D}}^{\dagger}(t)\mapsto\hat{\mathcal{D}}^{\dagger}{'}(t)=
 \hat{\mathcal{D}}^{\dagger}(t)+\frac{i}{4}(1-\alpha)\hat{\mathcal{B}}^{\dagger}(t)
\end{equation}
to allow us to rewrite the total Hamiltonian as
\begin{equation}
 \hat{H}=
 ik\{(\hat{\mathcal{B}}^{\dagger}(t)\hat{\mathcal{D}}{'}(t)-
 \hat{\mathcal{D}}^{\dagger}{'}(t)\hat{\mathcal{B}}(t))+
 (\hat{b}^{\dagger}(t)\hat{d}(t)-\hat{d}^{\dagger}(t)\hat{b}(t))\}.
\label{eq:totham1}
\end{equation}
Although the Hamiltonian becomes simple, the above shifts in
$\hat{\mathcal{D}}(t)$ and $\hat{\mathcal{D}}^{\dagger}(t)$ also changes the
commutation relation of them to
\begin{equation}
 [\hat{\mathcal{D}}(t),\hat{\mathcal{D}}^{\dagger}(t)]\mapsto
 [\hat{\mathcal{D}}{'}(t),\hat{\mathcal{D}}^{\dagger}{'}(t)]=\frac{1-\alpha}{2}.
\end{equation}
This makes the construction of a representation complicated compared
with the one given below. We thus take the $\alpha$-dependent form the
Hamiltonian as well as the commutators in Eq.~\eqref{eq:galg0} and
Eq.~\eqref{eq:galg1}.

Observing the analogy in the algebra as well as in the structure of
Hamiltonian between those of the gauge field and ghost fermions, we now
construct the basis of the vector space on which the operators are
represented. It will be easy to see that the operator
$\hat{H}_{\mathrm{G}}{'}$ in the Hamiltonian $\hat{H}_{\mathrm{G}}$,
expressed here by Schr\"odinger operators, has eigenvectors defined by
\begin{equation}
 \vert[n_{2}n_{1}]\rangle=\frac{1}{\sqrt{n_{1}!n_{2}!}}
 (\hat{\mathcal{D}}^{\dagger})^{n_{2}}
 (\hat{\mathcal{B}}^{\dagger})^{n_{1}}
 \vert0\rangle,\
 \vert[00]\rangle=\vert0\rangle,\quad
 n_{1},\
 n_{2}=0,\,1,\,2,\,\dots
\end{equation}
to satisfy
\begin{equation}
 \hat{H}_{\mathrm{G}}{'}\vert[n_{2}n_{1}]\rangle=
 k(n_{1}+n_{2}+1)\vert[n_{2}n_{1}]\rangle.
\end{equation}
Since taking hermitian conjugate changes the place of
$i\hat{\mathcal{B}}^{\dagger}\hat{\mathcal{D}}$ and
$-i\hat{\mathcal{D}}^{\dagger}\hat{\mathcal{B}}$ to each other, their right
eigenvectors will be also brought to the left ones of the other.
Hence, for $n_{1},\ n_{2}=0,\,1,\,2,\,\dots$,
\begin{equation}
 \langle[n_{2}n_{1}]\vert \hat{H}_{\mathrm{G}}{'}=
 k(n_{1}+n_{2}+1)\langle[n_{2}n_{1}]\vert,\
 \langle[n_{2}n_{1}]\vert=
 \frac{1}{\sqrt{n_{1}!n_{2}!}}\langle0\vert
 \hat{\mathcal{B}}^{n_{1}}
 \hat{\mathcal{D}}^{n_{2}}.
\end{equation}
An inner product of these eigenvectors is given by
\begin{equation}
 \langle[n_{2}n_{1}]\vert[n_{2}{'}n_{1}{'}]\rangle=
 i^{n_{1}-n_{2}}\delta_{n_{2}n_{1}{'}}\delta_{n_{1}n_{2}{'}}.
\label{eq:inprup0}
\end{equation}
This determines the metric structure of the vector space under
consideration. Then, to be consistent with this metric, we can construct
an expression for the resolution of unity:
\begin{equation}
 \sum_{n_{1},n_{2}=0}^{\infty}\vert[n_{2}n_{1}]\rangle
 i^{n_{2}-n_{1}}
 \langle[n_{1}n_{2}]\vert=1.
\label{eq:identity1a}
\end{equation}
Thus we have found that the vector space for representation of quantum
theory of our model is again equipped with indefinite metric(See
Eq.~\eqref{eq:identity0}). Similar to the fermionic case, the inner
product given above is brought to
\begin{equation}
 \langle\underline{[n_{2}n_{1}]}\vert[n_{1}{'}n_{2}{'}]\rangle=
 \delta_{n_{1}n_{1}{'}}\delta_{n_{2}n_{2}{'}}
\label{eq:inprup1}
\end{equation}
by introducing a conjugate $\langle\underline{[n_{2}n_{1}]}\vert$ for
$\vert[n_{2}n_{1}]\rangle$ defined by
\begin{equation}
 \vert[n_{2}n_{1}]\rangle\leftrightarrow
 \langle\underline{[n_{2}n_{1}]}\vert=
 \langle[n_{2}n_{1}]\vert\hat{\eta}_{\mathrm{G}},\
 \hat{\eta}_{\mathrm{G}}=
 e^{\pi i(\hat{\mathcal{B}}^{\dagger}+i\hat{\mathcal{D}}^{\dagger})
 (\hat{\mathcal{B}}-i\hat{\mathcal{D}})/2}\quad
 (\hat{\eta}^{-1}_{\mathrm{G}}=
 \hat{\eta}^{\dagger}_{\mathrm{G}}=\hat{\eta}_{\mathrm{G}}).
\end{equation}
The conjugation introduced above generates the following
transformation on operators:
\begin{equation}
 (\hat{\mathcal{B}},\hat{\mathcal{D}})\mapsto
 \hat{\eta}_{\mathrm{G}}^{\dagger}(\hat{\mathcal{B}},\hat{\mathcal{D}})
 \hat{\eta}_{\mathrm{G}}=
 (i\hat{\mathcal{D}},-i\hat{\mathcal{B}})
\end{equation}
together with their hermitian conjugations. In terms of the dual vectors
defined above, we can rewrite the identity operator as
\begin{equation}
 \sum_{n_{1},n_{2}=0}^{\infty}\vert[n_{2}n_{1}]\rangle
 \langle\underline{[n_{1}n_{2}]}\vert=1.
\label{eq:identity1}
\end{equation}

The analogy between ghost fermions and unphysical components of the
gauge field still continues to bring us the following definition of a
coherent state: 
\begin{equation}
 \vert\bbox{z}\rangle=
 e^{i(\hat{\mathcal{B}}^{\dagger}z_{2}-\hat{\mathcal{D}}^{\dagger}z_{1})}
 \vert0\rangle,\quad
 z_{1},\,z_{2}\in\bbox{C}.
\end{equation}
It will be straightforward to see
\begin{equation}
 \hat{\mathcal{B}}\vert\bbox{z}\rangle=z_{1}\vert\bbox{z}\rangle,\
 \hat{\mathcal{D}}\vert\bbox{z}\rangle=z_{2}\vert\bbox{z}\rangle,
\end{equation}
together with their hermitian conjugates. The inner product of these
coherent states will be given by
\begin{equation}
 \langle\bbox{z}^{*}\vert\bbox{z}{'}\rangle=
 e^{i(z_{1}^{*}z_{2}{'}-z_{2}^{*}z_{1}{'})},\
 \langle\bbox{z}^{*}\vert=(\vert\bbox{z}\rangle)^{\dagger}.
\end{equation}
It would be beautiful if we were able to give some meaning to the
integral
\begin{equation}
 \int\!\left(\frac{dz\,dz^{*}}{\pi}\right)^{2}\,
 e^{-i(z_{1}^{*}z_{2}-z_{2}^{*}z_{1})}
 \vert\bbox{z}\rangle
 \langle\bbox{z}^{*}\vert,\
 (dz\,dz^{*})^{2}=d\Re(z_{1})\,d\Im(z_{1})\,d\Re(z_{2})\,d\Im(z_{2})
\end{equation}
as the analogue of the resolution of unity for ghost fermions given by
Eq.~\eqref{eq:coh0}. A possible way to make the above integral
well-defined may be treating $-iz_{2}^{*}$ and $iz_{1}^{*}$ as complex
conjugate to $z_{1}$ and $z_{2}$, respectively. But it is actually
equivalent to replacing $\langle\bbox{z}^{*}\vert$ by
\begin{equation}
 \langle\underline{\bbox{z}}\vert=\langle0\vert
 e^{z_{1}^{*}\hat{\mathcal{B}}+z_{2}^{*}\hat{\mathcal{D}}}=
 \langle\bbox{z}^{*}\vert\hat{\eta}_{\mathrm{G}}
\end{equation}
to yield
\begin{equation}
 \int\!\left(\frac{dz\,dz^{*}}{\pi}\right)^{2}\,
 e^{-\bbox{z}^{\dagger}\bbox{z}}
 \vert\bbox{z}\rangle
 \langle\underline{\bbox{z}}\vert=1.
\label{eq:abcoh0}
\end{equation}
The analogy thus terminates here; we find only one formula of
resolution of unity in terms of coherent states for the case of the
gauge field while two formulas were possible for ghost fermions. The
reason for such difference is evident. In integrations with respect to
Grassmann numbers, we can treat $\xi$ and its conjugate entirely
independent to allow us the change of variables such that
$(-i\xi_{2}^{*},i\xi_{1}^{*})\mapsto(\xi_{1}^{*},\xi_{2}^{*})$ 
but this is not allowed for usual $c$-numbers.
It will be, therefore, natural that we
cannot find complete resemblance between bosonic and fermionic
degrees. Rather, we should be surprised at the existence of a formula
like Eq.~\eqref{eq:abcoh0} as an analogue of Eq.~\eqref{eq:coh1}.

\subsection{Coherent state path integral for unphysical degrees of the gauge field}
\label{ssec:cspi4gauge}
Acquiring the basic ingredient we now formulate a path integral for the
Hamiltonian $\hat{H}_{\mathrm{G}}$ in terms of the coherent
state. First, consider an infinitesimal version of the Euclidean kernel
defined by
\begin{equation}
 \langle\underline{\bbox{z}(j)}\vert
 \left(1-\epsilon \hat{H}_{\mathrm{G}}\right)
 \vert\bbox{z}(j-1)\rangle=
 \exp\left\{\bbox{z}^{\dagger}(j)(1-\epsilon kh)\bbox{z}(j-1)-
 \epsilon k\right\},\
\end{equation}
where $\epsilon=\beta/n$ and
\begin{equation}
 h=\begin{pmatrix}
    1&0\\
    -i(1-\alpha)/2&1
   \end{pmatrix}.
\end{equation}
Then repeated convolution of these infinitesimal kernels will bring us a
discretized path integral
\begin{multline}
 \langle\underline{\bbox{z}_{F}}\vert e^{-\beta \hat{H}_{\mathrm{G}}}
 \vert\bbox{z}_{I}\rangle=e^{-\beta k}
 \lim\limits_{n\to\infty}
 \int\!\prod_{i=1}^{n-1}\left(\frac{dz(i)\,dz^{*}(i)}{\pi}\right)^{2}\\
 \times
 \exp\left\{-\sum_{j=1}^{n-1}\bbox{z}^{\dagger}(j)\bbox{z}(j)
 +\sum_{j=1}^{n}\bbox{z}^{\dagger}(j)(1-\epsilon kh)\bbox{z}(j-1)\right\}.
\end{multline}
This can be evaluated in a straightforward way to be
\begin{equation}
 \langle\underline{\bbox{z}_{F}}\vert e^{-\beta \hat{H}_{\mathrm{G}}}
 \vert\bbox{z}_{I}\rangle=
 \exp\left\{-\beta k+
 e^{-\beta k}\bbox{z}^{\dagger}_{F}\gamma(\beta)\bbox{z}_{I}\right\},\
\label{eq:abkern0}
\end{equation}
where
\begin{equation}
  \gamma(t)=\begin{pmatrix}
    1&0\\
    -i(1-\alpha)kt/2&1
   \end{pmatrix}.
\label{eq:dipole-mat}
\end{equation}
Combining this result with Eq.~\eqref{eq:kern1} for ghost fermions, we
obtain for the total Hamiltonian
$\hat{H}=\hat{H}_{\mathrm{G}}+\hat{H}_{\mathrm{FP}}$
\begin{equation}
 \langle\underline{\bbox{z}_{F},\bbox{\xi}_{F}}\vert e^{-\beta \hat{H}}
 \vert\bbox{z}_{I},\bbox{\xi}_{I}\rangle=
 \exp\left\{
 e^{-\beta k}\bbox{z}^{\dagger}_{F}\gamma(\beta)\bbox{z}_{I}+
 e^{-\beta k}\bbox{\xi}^{\dagger}_{F}\bbox{\xi}_{I}
 \right\},
\label{eq:abkern1}
\end{equation}
where $\vert\bbox{z},\bbox{\xi}\rangle$ designates a direct product
given by
\begin{equation}
 \vert\bbox{z},\bbox{\xi}\rangle=
 \vert\bbox{z}\rangle\otimes\vert\bbox{\xi}\rangle.
\end{equation}
Since $\gamma(t)$ becomes unity for $\alpha=1$, we can see the complete
correspondence between FP ghosts and unphysical degrees of the gauge
field in the Feynman gauge. Indeed there exists a trivial symmetry, for
$\alpha=1$, in the total Hamiltonian given by Eq.~\eqref{eq:totham} under
exchanging the corresponding degrees between bosonic and fermionic
part. It must be, however, rather artificial that the symmetry is
restricted only to the Feynman gauge because a Feynman kernel is a gauge
dependent object. If we calculate a trace of time evolution operator to
remove the gauge dependence, we will find
\begin{equation}
\label{eq:trace0}
 \Tr(e^{-\beta H})=\int\left(\frac{dz\,dz^{*}}{\pi}\right)^{2}\,
 (d\xi\,d\xi^{*})^{2}\,
 e^{-\bbox{z}^{\dagger}\bbox{z}-\bbox{\xi}^{\dagger}\bbox{\xi}}
 \langle\underline{\bbox{z},\bbox{\xi}}\vert e^{-\beta \hat{H}}
 \vert\bbox{z},\bbox{\xi}\rangle
 =1
\end{equation}
regardless of the value of gauge parameter $\alpha$. Here we have
adopted the {\em periodic}\/ boundary condition\cite{HK1} for taking a
trace of the time evolution operator. It will be interesting to see that
we can generalize the periodic boundary condition for a trace formula to
a {\em twisted}\/ boundary condition in the following way
\begin{equation}
\label{eq:trace1}
 \Tr_{\theta}(e^{-\beta H})\equiv\int\left(\frac{dz\,dz^{*}}{\pi}\right)^{2}\,
 (d\xi\,d\xi^{*})^{2}\,
 e^{-\bbox{z}^{\dagger}\bbox{z}-\bbox{\xi}^{\dagger}\bbox{\xi}}
 \langle\underline{\bbox{z},\bbox{\xi}}\vert e^{-\beta \hat{H}}
 \vert e^{i\theta}\bbox{z},e^{i\theta}\bbox{\xi}\rangle
 =1
\end{equation}
without breaking the cancellation of determinants from both bosonic and
fermionic Gaussian integrals. Taking a trace over unphysical degrees will
not cause any effect on the physical partition function. Hence we are
free to make any choice for boundary conditions for unphysical
degrees. But the autonomy of unphysical degrees will be strongly
restricted by other reason, such as space-time symmetry for
example. Therefore if we require Lorentz covariance for the quantized
gauge field the unphysical degrees must obey the same boundary
condition with physical ones. This returns to the unique choice of the
boundary condition for FP ghosts because of the need for cancellation of
Gaussian determinants. Hence the periodic boundary condition even for
FP ghosts will be preferred in the calculation of a trace of a physical
quantity. 

Another point that should be remarked on our definition of the trace
formula is that, through the definitions of left eigenvectors of
creation operators, we have already included the metric structure of the
vector space under consideration. We have defined the coherent states to
yield the resolution of unity, taking the metric structures into
account, given by Eq.~\eqref{eq:identity} or Eq.~\eqref{eq:identity0} for
FP ghosts and by Eq.~\eqref{eq:identity1} for the gauge field. We may
define an operator given by
\begin{equation}
\label{eq:conjugate}
 \hat{\eta}=\hat{\eta}_{\mathrm{G}}\hat{\eta}_{\mathrm{FP}}=
 \hat{\eta}^{\dagger}=\exp\left[\frac{i\pi}{2}\left\{
 (\hat{\mathcal{B}}^{\dagger}+i\hat{\mathcal{D}}^{\dagger})
 (\hat{\mathcal{B}}-i\hat{\mathcal{D}})+
 (\hat{b}^{\dagger}+i\hat{d}^{\dagger})(\hat{b}-i\hat{d})\right\}\right]
\end{equation}
to find
\begin{equation}
 \langle\underline{\bbox{z},\bbox{\xi}}\vert=
 (\vert\bbox{z},\bbox{\xi}\rangle)^{\dagger}
 \hat{\eta}=\langle\bbox{z}^{*},\bbox{\xi}^{*}\vert\hat{\eta}
\end{equation}
and also
\begin{equation}
 \hat{\eta}^{\dagger}(\hat{\mathcal{B}},\hat{\mathcal{D}},b,d)\hat{\eta}=
 (i\hat{\mathcal{D}},-i\hat{\mathcal{B}},id,-ib).
\end{equation}
Note here that $\hat{A}$ and $\hat{A}_{0}$ transform under the
action of $\hat{\eta}$ as
\begin{equation}
 \hat{\eta}^{\dagger}\hat{A}\hat{\eta}=\hat{A},\
 \hat{\eta}^{\dagger}\hat{A}_{0}\hat{\eta}=-\hat{A}_{0}.
\end{equation}
By virtue of this relation, $\hat{A}_{0}$
becomes an hermitian-like operator on this indefinite metric vector
space. Further, we see that multiplication of $\hat{\eta}_{\mathrm{G}}$
is just the conventional one for dealing with the negative-norm property
of $\hat{A}_{0}$ in covariant quantization for electromagnetic field in
Feynman gauge.~\cite{YT} \ 

Thus we understand that we need to multiply this operator to complete
the hermitian conjugation on the vector space we are working
with. We may call the operator defined by Eq.~\eqref{eq:conjugate} as
conjugating operator hereafter. Hermiticity of operators must also be
defined to be consistent against this hermitian conjugation. It will be
then clear that left eigenvectors of creation operators are indeed the
hermitian conjugate to the right ones of annihilation operators.

\section{Generating functional and effective action}
\label{sec:gfefa}
We will find the effective actions for FP ghosts and unphysical degrees
of the gauge field by making use of coherent state path integral in this
section. To this aim we first consider the generating functionals for
these subsystems separately. Since the model we are working with is a toy
model of the free gauge field, the effective action found in this analysis
should have some analogue in the tree level calculation of usual
formulation of the conventional path integral for the gauge field and
hence expected to be trivial. However, it will be significant to confirm
such fundamental aspects of the method under development to make it
reliable.

To begin with, for a finite imaginary time $\beta=t_{F}-t_{I}$, we
consider a Feynman kernel of FP ghosts under the influence of external
Grassmann fields $\bbox{\eta}^{\dagger}(t)$ and $\bbox{\eta}(t)$ defined by
\begin{equation}
\begin{aligned}
 &K_{(\mathrm{ex})}(\bbox{\xi}_{F},t_{F};\bbox{\xi}_{I},t_{I})\\
 =&
 \langle\underline{\bbox{\xi}_{F};t_{F}}\vert
 \mathrm{T}\exp\left[-\int_{t_{1}}^{t_{2}}dt\,\left\{
 \bbox{\eta}^{\dagger}(t)
 \begin{pmatrix}
 \hat{b}(t)\\\hat{d}(t)
 \end{pmatrix}
 +(\hat{b}^{\dagger}(t),\hat{d}^{\dagger}(t))\bbox{\eta}(t)\right\}\right]
 \vert\bbox{\xi}_{I};t_{I}\rangle
\end{aligned}
\end{equation}
in which $\hat{b}(t)$, $\hat{d}(t)$ and their conjugate designate Heisenberg
operators for ghost fermions and coherent states are the left and right
eigenvectors of the Heisenberg operators at $t=t_{F}$ and $t=t_{I}$. The
lower and upper limits of the integration in the exponent defined by
T-product should be assumed to satisfy $t_{F}>t_{2}>t_{1}>t_{I}$.

Dividing $\beta$ into $n$ pieces and making use of the resolution of
unity Eq.~\eqref{eq:coh1} repeatedly, we obtain a path integral
\begin{multline}
 K_{(\mathrm{ex})}(\bbox{\xi}_{F},t_{F};\bbox{\xi}_{I},t_{I})=
 \lim\limits_{n\to\infty}\int\prod_{i=1}^{n-1}(d\xi(i)\,d\xi^{*}(i))^{2}\\
 \times
 \exp\left\{{-\sum_{j=1}^{n-1}
 \bbox{\xi}^{\dagger}(j)\bbox{\xi}(j)+
 (1-\epsilon k)\sum_{j=1}^{n}
 \bbox{\xi}^{\dagger}(j)\bbox{\xi}(j-1)}\right\}\\
 \times
 \exp\left\{{-\epsilon\sum_{j=1}^{n}
 \bbox{\eta}^{\dagger}(j)\bbox{\xi}(j-1)-
 \bbox{\xi}^{\dagger}(j)\sigma_{2}\bbox{\eta}(j)}\right\},
\end{multline}
where $\sigma_{2}$ is the Pauli matrix. Here and in the following we
will drop the constant $\mp k$ in the Hamiltonians for FP ghosts and
unphysical degrees of the gauge field because they cancel each other as
was already stated. The Gaussian integration can be performed easily and
we can take continuum limit to find
\begin{multline}
 K_{(\mathrm{ex})}(\bbox{\xi}_{F},t_{F};\bbox{\xi}_{I},t_{I})=
 \exp\left\{e^{-k(t_{F}-t_{I})\bbox{\xi}_{F}^{\dagger}\bbox{\xi}_{I}}\right\}\\
 \times
 \exp\left\{{\int_{t_{1}}^{t_{2}}\!\!dt\,
 e^{-k(t_{F}-t)}\bbox{\xi}^{\dagger}_{F}\sigma_{2}\bbox{\eta}(t)-
 e^{-k(t-t_{I})}\bbox{\eta}^{\dagger}(t)\bbox{\xi}_{I}}\right\}\\
 \times
 \exp\left\{{-\int_{t_{1}}^{t_{2}}\!\!dt\,dt{'}\,
 \theta(t-t{'})e^{-k(t-t{'})}
 \bbox{\eta}^{\dagger}(t)\sigma_{2}\bbox{\eta}(t{'})}\right\},
\end{multline}
where $\theta(t)$ is the step function.
By taking limits $t_{F},-t_{I}\to\infty$ and then
$-t_{1},t_{2}\to\infty$ in this order, we can find a generating
functional
\begin{equation}
\begin{aligned}
 & Z^{(\mathrm{FP})}[\bbox{\eta},\bbox{\eta}^{\dagger}]=
 e^{-W^{(\mathrm{FP})}[\bbox{\eta},\bbox{\eta}^{\dagger}]}\\
 =&\langle\{00\}\vert
 \mathrm{T}\exp\left[-\int_{-\infty}^{\infty}dt\,\left\{
 \bbox{\eta}^{\dagger}(t)
 \begin{pmatrix}
 \hat{b}(t)\\\hat{d}(t)
 \end{pmatrix}
 +(\hat{b}^{\dagger}(t),\hat{d}^{\dagger}(t))\bbox{\eta}(t)\right\}\right]
 \vert\{00\}\rangle\\
 =&
 \exp\left\{{-\int_{-\infty}^{\infty}\!\!dt\,dt{'}\,
 \theta(t-t{'})e^{-k(t-t{'})}
 \bbox{\eta}^{\dagger}(t)\sigma_{2}\bbox{\eta}(t{'})}\right\}.
\end{aligned}
\end{equation}
From the generating functional, we can read the propagator of FP ghosts
\begin{equation}
 \langle\{00\}\vert\mathrm{T}
 \begin{pmatrix}
 \hat{b}(t)\\
 \hat{d}(t)
 \end{pmatrix}
 \begin{pmatrix}
  \hat{b}^{\dagger}(t{'}),\,\hat{d}^{\dagger}(t{'})
 \end{pmatrix}
 \vert\{00\}\rangle=
 -\theta(t-t{'})e^{-k(t-t{'})}
 \begin{pmatrix}
  0&-i\\
  i&0
 \end{pmatrix}.
\end{equation}
If we put
\begin{equation}
 \bbox{\eta}^{\dagger}(t)=\frac{1}{\sqrt{2k}}
 \begin{pmatrix}
  \bar{\psi}(t),\,-\psi(t)
 \end{pmatrix},\
 \bbox{\eta}(t)=\frac{1}{\sqrt{2k}}
 \begin{pmatrix}
  -\bar{\psi}(t)\\
 \psi(t)
 \end{pmatrix}
\end{equation}
to render the external sources couple to $\hat{c}(t)$ and
$\hat{\bar{c}}(t)$, we will find a generating functional of connected
Green's function given by
\begin{equation}
\label{eq:genfunc01}
 W^{(\mathrm{FP})}[\bar{\psi},\psi]=
 -\frac{1}{2}\int_{-\infty}^{\infty}\!\!dt\,dt{'}\,
 \frac{1}{i}\Delta_{F}(t-t{'})
 \left\{\bar{\psi}(t)\psi(t{'})-\psi(t)\bar{\psi}(t{'})\right\}
\end{equation}
in which $\Delta_{F}(t)$ being the Feynman propagator
\begin{equation}
 \Delta_{F}(t)=\frac{1}{2\pi}\int_{-\infty}^{\infty}\!\!dp_{0}\,
 e^{-ip_{0}t}\frac{1}{p^{2}},\quad
 p^{2}=p_{0}^{2}+k^{2}.
\end{equation}
We may write expectation values of $c(t)$ and $\bar{c}(t)$ under the
external sources as
\begin{equation}
 c(t)=\delta W^{(\mathrm{FP})}[\bar{\psi},\psi]/\delta\bar{\psi}(t),\
 \bar{c}(t)=\delta W^{(\mathrm{FP})}[\bar{\psi},\psi]/\delta\psi(t),
\end{equation}
where functional derivatives are defined by right action, to find that a
Euclidean effective action of FP ghosts in the tree level is given by
\begin{equation}
\label{eq:eucea01}
 \varGamma_{E}^{(\mathrm{FP})}[\bar{c},c]=
 c\cdot\bar{\psi}+\bar{c}\cdot\psi-W^{(\mathrm{FP})}=
 i\int_{-\infty}^{\infty}\!\!dt\,
 \left\{\dot{\bar{c}}(t)\dot{c}(t)+k^{2}\bar{c}(t)c(t)\right\}.
\end{equation}
Here and in the following we may often use a notation
\begin{equation}
 f\cdot g\equiv\int_{-\infty}^{\infty}\!\!dt\,f(t)g(t).
\end{equation}
The Euclidean effective action in Eq.~\eqref{eq:eucea01} is translated
into the corresponding one for a Minkowski(real) time $t$ by inverse
Wick rotation, $t\to it$;
\begin{equation}
 \varGamma_{E}^{(\mathrm{FP})}[\bar{c},c]\xrightarrow{\quad t\to it\quad}
 i\varGamma^{(\mathrm{FP})}[\bar{c},c]=\int_{-\infty}^{\infty}\!\!dt\,
 \left\{\dot{\bar{c}}(t)\dot{c}(t)-k^{2}\bar{c}(t)c(t)\right\}
\end{equation}
which is nothing but the classical action(times $i$) for the
FP ghosts. We thus see the validity of our method of constructing
Euclidean path integral in terms of coherent states for ghost fermions.

In the same way we can calculate a generating functional for unphysical
degrees of the gauge field by means of coherent state path integral. The
process of calculation is quite familiar one hence we omit the detail
but simply list some results below.
The generating functional with Feynman's boundary condition is defined
and given by
\begin{equation}
\begin{aligned}
 & Z^{(\mathrm{G})}[\bbox{j},\bbox{j}^{\dagger}]=
 e^{-W^{(\mathrm{G})}[\bbox{j},\bbox{j}^{\dagger}]}\\
 =&\langle[00]\vert
 \mathrm{T}\exp\left[-\int_{-\infty}^{\infty}dt\,\left\{
 \bbox{j}^{\dagger}(t)
 \begin{pmatrix}
 \hat{\mathcal{B}}(t)\\ \hat{\mathcal{D}}(t)
 \end{pmatrix}
 +(\hat{\mathcal{B}}^{\dagger}(t),\hat{\mathcal{D}}^{\dagger}(t))
 \bbox{j}(t)\right\}\right]
 \vert[00]\rangle\\
 =&
 \exp\left\{{-\int_{-\infty}^{\infty}\!\!dt\,dt{'}\,
 \theta(t-t{'})e^{-k(t-t{'})}
 \bbox{j}^{\dagger}(t)\gamma(t-t{'})\sigma_{2}\bbox{j}(t{'})}\right\},
\end{aligned}
\end{equation}
where $\gamma(t)$ is the one defined in Eq.~\eqref{eq:dipole-mat}. From
the generating functional we read the propagator
\begin{equation}
 \langle[00]\vert\mathrm{T}
 \begin{pmatrix}
 \hat{\mathcal{B}}(t)\\
 \hat{\mathcal{D}}(t)
 \end{pmatrix}
 \begin{pmatrix}
  \hat{\mathcal{B}}^{\dagger}(t{'}),\,\hat{\mathcal{D}}^{\dagger}(t{'})
 \end{pmatrix}
 \vert[00]\rangle=
 -\theta(t-t{'})e^{-k(t-t{'})}
 \begin{pmatrix}
  0&-i\\
  i&(1-\alpha)k(t-t{'})/2
 \end{pmatrix}.
\end{equation}
The existence of a component that is linear in $t-t{'}$ in the
propagator clearly exhibits the effect of dipole ghost in the unphysical
degrees of the gauge field.

By putting
\begin{equation}
\begin{gathered}
 \bbox{j}^{\dagger}(t)=\frac{1}{\sqrt{2k}}
 \begin{pmatrix}
 -\dfrac{1+\alpha}{4k}(J(t)+ikJ_{0}(t))+kJ_{B}(t),\,
 -\dfrac{i}{k}(J(t)-ikJ_{0}(t))
 \end{pmatrix},\\
 \bbox{j}(t)=\frac{1}{\sqrt{2k}}
 \begin{pmatrix}
 -\dfrac{1+\alpha}{4k}(J(t)-ikJ_{0}(t))+kJ_{B}(t)\\
 \dfrac{i}{k}(J(t)+ikJ_{0}(t))
 \end{pmatrix}
\end{gathered}
\end{equation}
to change the source term to
$J\cdot \hat{A}+J_{0}\cdot\hat{A}_{0}+J_{B}\cdot\hat{B}$,
we find a generating functional of connected Green's functions
for unphysical components of the gauge field given by
\begin{equation}
\label{eq:genfunc02}
 W^{(\mathrm{G})}[J,J_{0},J_{B}]=
 -\frac{1}{2}\int_{-\infty}^{\infty}\!\!dt\,dt{'}\,
 \bbox{J}^{T}(t)D_{F}(t-t{'})\bbox{J}(t{'}),\quad
 \bbox{J}^{T}(t)=(J(t),J_{0}(t),J_{B}(t)),
\end{equation}
in which the Feynman propagator $D_{F}(t)$ for the gauge field is
defined by
\begin{equation}
\begin{gathered}
 D_{F}(t)=\frac{1}{2\pi}\int_{-\infty}^{\infty}\!\!dp_{0}\,
 e^{-ip_{0}t}\tilde{D}_{F}(p_{0}),\\
 \tilde{D}_{F}(p_{0})=\frac{1}{p^{2}}
 \begin{pmatrix}
  \dfrac{1}{k^{2}}\left\{1-(1-\alpha)\dfrac{k^{2}}{p^{2}}\right\}&
  (1-\alpha)\dfrac{p_{0}}{p^{2}}&-1\\
  -(1-\alpha)\dfrac{p_{0}}{p^{2}}&
  -1+(1-\alpha)\dfrac{p_{0}^{2}}{p^{2}}&-p_{0}\\
  -1&p_{0}&0
 \end{pmatrix},
\end{gathered}
\end{equation}
where $p^{2}=p_{0}^{2}+k^{2}$.
The effective action for unphysical degrees of the gauge field is then
found to be 
\begin{equation}
 \begin{aligned}
 &\varGamma_{E}^{(\mathrm{G})}[A,A_{0},B]=
 A\cdot J+A_{0}\cdot J_{0}+B\cdot J_{B}-W^{(\mathrm{G})}\\
 =&\frac{1}{2}\int_{-\infty}^{\infty}\!\!dt\,
 \left\{k^{2}(i\dot{A}(t)-A_{0}(t))^{2}-2i\dot{B}(t)A_{0}(t)+
 2k^{2}A(t)B(t)+\alpha B^{2}(t)\right\},
\end{aligned}
\label{eq:efaceuc}
\end{equation}
where
\begin{equation}
 A(t)=\frac{\delta W^{(\mathrm{G})}}{\delta J(t)},\
 A_{0}(t)=\frac{\delta W^{(\mathrm{G})}}{\delta J_{0}(t)},\
 B(t)=\frac{\delta W^{(\mathrm{G})}}{\delta J_{B}(t)}.
\end{equation}
Again by inverse Wick rotation, we obtain the classical action for
unphysical degrees of the gauge field
\begin{equation}
 \begin{aligned}
 &\varGamma_{E}^{(\mathrm{G})}\xrightarrow{\quad t\to it\quad}
 i\varGamma^{(\mathrm{G})}\\
 =&i\int_{-\infty}^{\infty}\!\!dt\,\frac{1}{2}
 \left\{k^{2}(\dot{A}(t)-A_{0}(t))^{2}-2\dot{B}(t)A_{0}(t)+
 2k^{2}A(t)B(t)+\alpha B^{2}(t)\right\}
\end{aligned}
\label{eq:efacmin}
\end{equation}
as an effective action of zeroth order calculation in the perturbative
expansion. We thus confirmed that Euclidean path integral for generating
functional of FP ghosts and unphysical components of the gauge field in
terms of coherent states constructed in preceding sections can reproduce
classical actions for these variables in the leading order of
perturbation theory. Hence we may regard our method to be reliable.

\section{BRS quartet and Kugo-Ojima projection}
\label{sec:kugoojima}
In this section we will classify the state vectors appear in the vector
space of the representation for the quantum system under
consideration. Then we will find an explicit form of the Kugo-Ojima
projection in term of the field variables.
From the BRS invariance of the Lagrangian there follows a conserved
Noether charge(BRS charge) $Q_{B}$ given by
\begin{equation}
 \label{eq:brscharge}
 Q_{B}=\hat{P}(t)\hat{c}(t)-i\hat{B}(t)\hat{p}_{\bar{c}}(t)=
 -\dot{\hat{B}}(t)\hat{c}(t)+\hat{B}(t)\dot{\hat{c}}(t).
\end{equation}
This can be expressed in terms of the creation and annihilation
operators as
\begin{equation}
 Q_{B}=k\hat{Q}_{B},\
 \hat{Q}_{B}=i(\hat{\mathcal{B}}\hat{b}^{\dagger}-
 \hat{\mathcal{B}}^{\dagger}\hat{b})
\end{equation}
where we have employed the Schr\"odinger picture. Among basis vectors
only the vacuum state
\begin{equation}
 \vert0\rangle=\vert[00]\rangle\otimes\vert\{00\}\rangle
\end{equation}
is classified as a BRS singlet because it has a positive norm and BRS
invariant. Other state vectors, given by
\begin{equation}
 \vert[m_{2}m_{1}]\{n_{2}n_{1}\}\rangle=
 \vert[m_{2}m_{1}]\rangle\otimes\vert\{n_{2}n_{1}\}\rangle,\
 m_{i}=0,\,1,\,\,2,\,\dots,\
 n_{i}=0,\,1,
\end{equation}
have zero-norm excepting the diagonal ones, specified by
\begin{equation}
 \vert[mm]\{nn\}\rangle
\end{equation}
if $m_{1}+m_{2}+n_{1}+n_{2}\ge1$. Though norms of basic vectors
diagonal both within bosonic and fermionic sectors themselves do not
vanish, they are arranged to have zero-norm in a pair wise manner within
a BRS quartet\cite{KO} as will be shown below.

For a given pair of $m$ and $n$($m,\,n=0,\,1,\,2,\,\dots$) we can
see the following cyclic sequence of finding BRS-quartet.
First, take $\vert[mn]\{10\}\rangle$ and make BRS transform by
multiplying $\hat{Q}_{B}$ to obtain a BRS-exact state vector: 
\begin{equation}
\label{eq:brscyc1}
 \begin{aligned}
  \hat{Q}_{B}\vert[mn]\{10\}\rangle=
  \sqrt{m}\vert[m-1n]\{11\}\rangle+
  \sqrt{n+1}\vert[mn+1]\{00\}\rangle.
 \end{aligned}
\end{equation}
It is then brought to its partner with respect to the inner product by a
multiplication of $\hat{\eta}$, defined by Eq.~\eqref{eq:conjugate},
\begin{equation}
\label{eq:brscyc2}
 \hat{\eta}\hat{Q}_{B}\vert[mn]\{10\}\rangle=
 i^{m-n-1}(-\sqrt{m}\vert[nm-1]\{11\}\rangle+
 \sqrt{n+1}\vert[n+1m]\{00\}\rangle).
\end{equation}
BRS transform again on this vector will produces another BRS-exact
vector given by
\begin{equation}\label{eq:brscyc3}
 \hat{Q}_{B}\hat{\eta}\hat{Q}_{B}\vert[mn]\{10\}\rangle=
 i^{m-n-1}(m+n+1)\vert[mn]\{10\}\rangle
\end{equation}
and finally conjugation again to find
\begin{equation}
\label{eq:brscyc4}
 \hat{\eta}\hat{Q}_{B}\hat{\eta}\hat{Q}_{B}\vert[mn]\{10\}\rangle=
 (m+n+1)\vert[mn]\{10\}\rangle
\end{equation}
and close the cycle. We thus obtain the following cyclic diagram
\begin{equation}
\label{eq:brscys}
 \begin{CD}
 \vert[mn]\{10\}\rangle @>\hat{Q}_{B}>> 
 \sqrt{m}\vert[m-1n]\{11\}\rangle+
 \sqrt{n+1}\vert[mn+1]\{00\}\rangle\\
 @A{\hat{\eta}}AA @VV{\hat{\eta}}V\\
 \vert[nm]\{01\}\rangle @<<\hat{Q}_{B}<
 -\sqrt{m}\vert[nm-1]\{11\}\rangle+
 \sqrt{n+1}\vert[n+1m]\{00\}\rangle
\end{CD}
\end{equation}
for a BRS quartet. From Eq.~\eqref{eq:brscyc4}, we are naturally
lead to define another quantity, though it is not conserved in general,
$\hat{Q}_{D}$ by
\begin{equation}
\label{eq:antibrs}
 \hat{Q}_{D}=\hat{\eta}\hat{Q}_{B}\hat{\eta}=
 i(\hat{\mathcal{D}}\hat{d}^{\dagger}-\hat{\mathcal{D}}^{\dagger}\hat{d})
\end{equation}
and to rewrite above BRS cyclic diagram as
\begin{equation}
\label{eq:abrscys}
 \begin{CD}
 \vert[mn]\{10\}\rangle @<\hat{Q}_{D}<< 
 \sqrt{m}\vert[m-1n]\{11\}\rangle+
 \sqrt{n+1}\vert[mn+1]\{00\}\rangle\\
 @V{\hat{\eta}}VV @AA{\hat{\eta}}A\\
 \vert[nm]\{01\}\rangle @>>\hat{Q}_{D}>
 -\sqrt{m}\vert[nm-1]\{11\}\rangle+
 \sqrt{n+1}\vert[n+1m]\{00\}\rangle
\end{CD}
\end{equation}
to be viewed as a cycle of {\em BRS-inversion} in a quartet. Here, the
meaning of BRS-inversion will be clear from
above diagrams. It will be interesting to see that BRS
variant members of a BRS quartet are exact under the BRS-inversion hence
zero-normed.

Since a member of BRS quartet has its partner with respect to the inner
product only within the same quartet, in which only two members are
physical states satisfying
\begin{equation}
 \hat{Q}_{B}\vert\text{Phys}\rangle=0,
\end{equation}
the BRS quartet spans a four dimensional subspace of the total vector
space. A projection to this subspace may be expressed as
\begin{equation}
 P_{mn}=\vert B_{mn}^{(0)}\rangle\langle A_{mn}^{(0)}\vert+
 \vert B_{mn}^{(+)}\rangle\langle A_{mn}^{(-)}\vert+
 \vert A_{mn}^{(0)}\rangle\langle B_{mn}^{(0)}\vert+
 \vert A_{mn}^{(-)}\rangle\langle B_{mn}^{(+)}\vert,
\end{equation}
where 
\begin{equation}
 \begin{aligned}
  \vert A_{mn}^{(-)}\rangle=&\vert[mn]\{10\}\rangle,\\
  \vert A_{mn}^{(0)}\rangle=&
  \frac{1}{\sqrt{m+n+1}}\hat{\eta}\hat{Q}_{B}\vert[mn]\{10\}\rangle,\\
  \vert B_{mn}^{(0)}\rangle=&\frac{1}{\sqrt{m+n+1}}
  \hat{Q}_{B}\vert[mn]\{10\}\rangle,\\
  \vert B_{mn}^{(+)}\rangle=&\frac{1}{m+n+1}
  \hat{Q}_{B}\hat{\eta}\hat{Q}_{B}\vert[mn]\{10\}\rangle.
 \end{aligned}
\end{equation}
The total vector space under consideration is then decomposed into a
direct sum of these subspaces in addition to the one dimensional really
physical subspace spanned by the vacuum state. Hence we recognize that a
sum of all $P_{mn}$ given above is expressed as
\begin{equation}
 \sum_{m,n=0}^{\infty}P_{mn}=1-\vert0\rangle\langle0\vert.
\end{equation}
If we partially sum $P_{mn}$ in the above entire summation by putting
$m+n+1=l$ with a positive integer $l$, we obtain a projection to a $4l$
dimensional subspace given by
\begin{equation}
\label{eq:koprj0}
 P^{(l)}=\sum_{m+n+1=l}P_{mn}.
\end{equation}
We thus obtain
\begin{equation}
 \sum_{l=0}^{\infty}P^{(l)}=1,\quad
 P^{(0)}\equiv\vert0\rangle\langle0\vert.
\end{equation}

Let us now give an explicit form of the projection $P^{(l)}$, which is
nothing but the Kugo-Ojima projection\cite{KO}, in terms of
the creation and annihilation operators introduced before. The
Kugo-Ojima projection $P^{(l)}$ for the system in consideration is given
by
\begin{equation}
\label{eq:kugoojimaprj}
 P^{(l)}=\frac{1}{2\pi}\int_{0}^{2\pi}\!\!d\theta\,
 \exp\left\{i\theta(\hat{N}-l)\right\},\quad
 \hat{N}=i\left\{(\hat{\mathcal{B}}^{\dagger}\hat{\mathcal{D}}-
 \hat{\mathcal{D}}^{\dagger}\hat{\mathcal{B}})+
 (\hat{b}^{\dagger}\hat{d}-\hat{d}^{\dagger}\hat{b})
 \right\}.
\end{equation}
The proof will be quite simple because $\hat{N}$ is the operator that
counts the total number of excitations in a state vector
$\vert[m_{2}m_{1}]\{n_{2}n_{1}\}\rangle$ so that
\begin{equation}
 \hat{N}\vert[m_{2}m_{1}]\{n_{2}n_{1}\}\rangle=
 (m_{1}+m_{2}+n_{1}+n_{2})\vert[m_{2}m_{1}]\{n_{2}n_{1}\}\rangle
\end{equation}
while $m+n+1$ being the common eigenvalue of $\hat{N}$ on all members of
a BRS quartet of which $\vert[mn]\{10\}\rangle$ being a member.
It will be worth noting that the operator $\hat{N}$ introduced above is
essentially the Hamiltonian for the total system given by
Eq.~\eqref{eq:totham}. In particular, when we employ the Feynman gauge by
choosing $\alpha=1$, the gauge parameter dependent part disappears from
the Hamiltonian to result in $\hat{H}=k\hat{N}$. It holds, however, that
two parts of the Hamiltonian Eq.~\eqref{eq:totham} are expressed as 
a BRS transform separately for any gauge parameter $\alpha$. Actually
and indeed surprisingly, the operator $\hat{N}$ can be expressed by the
anticommutator 
\begin{equation}
 \hat{N}=\{\hat{Q}_{B},\hat{Q}_{D}\}.
\end{equation}
Hence $\hat{N}$ itself is BRS exact and BRS-inversion exact
simultaneously. It is also true for the Hamiltonian for the Feynman
gauge. The fact that $\hat{N}$ can be expressed in a BRS exact form
explains the reason why Kugo-Ojima projection $P^{(l)}$ can be written
as an anticommutator with $R^{(l)}$ by explicit construction(See Eq.(3.29)
of Ref.~\cite{KO}).

An immediate application of the formula Eq.~\eqref{eq:kugoojimaprj} for
Kugo-Ojima projection will be found in the calculation of projection
inserted Feynman kernel defined by
\begin{equation}
 K^{(l)}(\bbox{z}_{F},\bbox{\xi}_{F};\bbox{z}_{I},\bbox{\xi}_{I};\beta)\equiv
 \langle\underline{\bbox{z}_{F},\bbox{\xi}_{F}}\vert
 e^{-\beta \hat{H}}P^{(l)}\vert\bbox{z}_{I},\bbox{\xi}_{I}\rangle.
\label{eq:kernprj0}
\end{equation}
By use of the formula Eq.~\eqref{eq:kugoojimaprj}, the calculation of
this kernel reduces to
\begin{equation}
\begin{aligned}
 K^{(l)}(\bbox{z}_{F},\bbox{\xi}_{F};\bbox{z}_{I},\bbox{\xi}_{I};\beta)=&
 \frac{1}{2\pi}\int_{0}^{2\pi}\!\!d\theta\,e^{-il\theta}
 \exp\left[
 e^{-\beta k+i\theta}\left\{\bbox{z}^{\dagger}_{F}\gamma(\beta)\bbox{z}_{I}+
 \bbox{\xi}^{\dagger}_{F}\bbox{\xi}_{I}\right\}
 \right] \\
 =&
 e^{-l\beta k}\left\{\bbox{z}^{\dagger}_{F}\gamma(\beta)\bbox{z}_{I}+
 \bbox{\xi}^{\dagger}_{F}\bbox{\xi}_{I}\right\}^{l},
\end{aligned}
\label{eq:kernprj1}
\end{equation}
because the existence of $e^{i\theta\hat{N}}$ causes changes to the path
integral only in the shift of $\beta k$ to $\beta k+i\theta$ excepting
the $\alpha$ dependent part.
If we apply the same technique to the calculation of trace formula
Eq.~\eqref{eq:trace0}, we will obtain
\begin{equation}
\begin{aligned}
 Z^{(l)}(\beta)=&
 \frac{1}{2\pi}\int_{0}^{2\pi}\!\!d\theta\,e^{-il\theta}
 \int\left(\frac{dz\,dz^{*}}{\pi}\right)^{2}\,
 (d\xi\,d\xi^{*})^{2}\,
 e^{-\bbox{z}^{\dagger}\bbox{z}-\bbox{\xi}^{\dagger}\bbox{\xi}}\\
 &\times
 \exp\left[
 e^{-\beta k+i\theta}\left\{\bbox{z}^{\dagger}\gamma(\beta)\bbox{z}+
 \bbox{\xi}^{\dagger}\bbox{\xi}\right\}
 \right] \\
 =&
 \frac{1}{2\pi}\int_{0}^{2\pi}\!\!d\theta\,e^{-il\theta}\\
 =&\delta_{l0}
 \end{aligned}
\label{eq:zprj1}
\end{equation}
to observe that only the vacuum state can contribute to the partition
function. On the other hand, we may perform the integration with respect
to $\theta$ ahead, that is equivalent to insert the projection
$P^{{(l)}}$ in the form given by Eq.~\eqref{eq:koprj0}, to find
\begin{equation}
 Z^{(l)}(\beta)=
 e^{-l\beta k}\int\left(\frac{dz\,dz^{*}}{\pi}\right)^{2}\,
 (d\xi\,d\xi^{*})^{2}\,
 e^{-\bbox{z}^{\dagger}\bbox{z}-\bbox{\xi}^{\dagger}\bbox{\xi}}
 \left\{\bbox{z}^{\dagger}\gamma(\beta)\bbox{z}+
 \bbox{\xi}^{\dagger}\bbox{\xi}\right\}^{l}.
\label{eq:zprj2}
\end{equation}
For this case $Z^{(l)}=0(l\ge1)$ must be checked term by term. This
clearly exhibits the usefulness of the formula
Eq.~\eqref{eq:kugoojimaprj}; The set of infinitely many identities in
Eq.~\eqref{eq:zprj1} as a consequence of the BRS invariance must be
checked directly for each $l=0,\,1,\,2,\,\dots$ if we do not have the
expression of $P^{(l)}$ in Eq.~\eqref{eq:kugoojimaprj}. Therefore it
should be stressed again that the knowledge of the explicit form of
Kugo-Ojima projection, that is having the formula given by 
Eq.~\eqref{eq:kugoojimaprj} in hand, is quite significant.

\section{Application to the free gauge field}
\label{sec:maxwell}
Having completed the thorough study of the toy model, let us consider
the application of our technique developed in preceding sections to
quantum theory of the abelian gauge field. We first rewrite the
Lagrangian for a free gauge field
\begin{equation}
 \mathcal{L}_{0}(x)=-\frac{1}{4}F_{\mu\nu}(x)F^{\mu\nu}(x),\
 F_{\mu\nu}(x)=\partial_{\mu}A_{\nu}(x)-\partial_{\nu}A_{\mu}(x)
\end{equation}
by parameterizing spatial components as
\begin{equation}
 \bbox{A}(x)=-\nabla A(x)+\bbox{A}_{\mathrm{T}}(x),\
 \nabla\cdot\bbox{A}_{\mathrm{T}}(x)=0
\end{equation}
to find
\begin{equation}
 \mathcal{L}_{0}(x)=\frac{1}{2}\left\{\nabla(\dot{A}(x)-A_{0}(x))\right\}^{2}+
 \frac{1}{2}\dot{\bbox{A}}_{\mathrm{T}}^{2}(x)-
 \frac{1}{2}(\nabla\times\bbox{A}_{\mathrm{T}}(x))^{2}.
\end{equation}
Clearly the second term together with the third one in this Lagrangian
describes two physical degrees of the abelian gauge field in the Coulomb
gauge and its properties including quantum theory on a Fock space with
positive definite metric are quite well-known. Therefore our target in
this section is the first term. Let us write it as
\begin{equation}
 \mathcal{L}_{0}{'}(x)=
 \frac{1}{2}\left\{\nabla(\dot{A}(x)-A_{0}(x))\right\}^{2}
\end{equation}
and make an investigation on the quantization of this Lagrangian with a
covariant gauge condition
\begin{equation}
 \dot{A}_{0}(x)-\nabla^{2}A(x)+\alpha B(x)=0
\end{equation}
in which Nakanishi-Lautrup\cite{NL} field $B(x)$ has been
introduced. Although we are dealing with this covariant type gauge
condition, the space-time covariance has already been lost by the
decomposition of the original Lagrangian into two non-covariant
parts. The covariance will, however, be restored if we recombine them
together after quantization by explicit calculation of an effective action
for entire system. Hence we discard the absence of space-time covariance
in the subsystem we are working with for a while.

We introduce gauge fixing and FP ghosts according to the BRS formalism
by adding
\begin{equation}
 \mathcal{L}_{\mathrm{GF}+\mathrm{FP}}(x)=i\bbox{\delta}_{B}\left\{\bar{c}(x)
 (\dot{A}_{0}(x)-\nabla^{2}A(x)+\alpha B^{2}(x)/2)
 \right\}
\end{equation}
to $\mathcal{L}{'}(x)$ to find
\begin{equation}
\begin{gathered}
 \mathcal{L}(x)=\mathcal{L}_{\mathrm{G}}(x)+\mathcal{L}_{\mathrm{FP}}(x),\\
 \mathcal{L}_{\mathrm{G}}(x)=\mathcal{L}_{0}{'}(x)-
 \dot{B}(x)A_{0}(x)+\nabla B(x)\cdot\nabla
 A(x)+\frac{\alpha}{2}B^{2}(x),\\
 \mathcal{L}_{\mathrm{FP}}(x)=-i\partial_{\mu}\bar{c}(x)\partial^{\mu}c(x).
\end{gathered}
\end{equation}
From variation of this Lagrangian we can immediately obtain equations of
motion for field variables. As is naturally expected, unphysical degrees
of the gauge field $A(x)$ and $A_{0}(x)$ satisfy
\begin{align}
 \square A(x)=&(1-\alpha)B(x),\\
 \square A_{0}(x)=&(1-\alpha)\dot{B}(x),
\end{align}
while $B(x)$, $c(x)$ and $\bar{c}(x)$ are obeying massless free field
equations
\begin{equation}
 \square B(x)=
 \square c(x)=
 \square \bar{c}(x)=0.
\end{equation}
We thus quantize FP ghost and anti-ghost as free fields in a quite
similar manner as has been done for $\hat{c}(t)$ and $\hat{\bar{c}}(t)$
in the subsection \ref{ssec:ghosts} for each Fourier components in
\begin{equation}
\begin{aligned}
 \hat{c}(x)=&\int\!\!\frac{d^{3}k}{\sqrt{(2\pi)^{3}2\vert\bbox{k}\vert}}\,
 \left\{\hat{b}(\bbox{k})e^{-ikx}+
 \hat{b}^{\dagger}(\bbox{k})e^{-ikx}\right\},\\
 \hat{\bar{c}}(x)=
 &\int\!\!\frac{d^{3}k}{\sqrt{(2\pi)^{3}2\vert\bbox{k}\vert}}\,
 \left\{\hat{d}(\bbox{k})
 e^{-ikx}+\hat{d}^{\dagger}(\bbox{k})e^{-ikx}\right\},
\end{aligned}
\end{equation}
with the following anticommutation relations
\begin{equation}
 \{\hat{b}(\bbox{p}),\hat{d}^{\dagger}(\bbox{q})\}=
 i\delta^{3}(\bbox{p}-\bbox{q}),\
 \{\hat{b}^{\dagger}(\bbox{p}),\hat{d}(\bbox{q})\}=
 -i\delta^{3}(\bbox{p}-\bbox{q}),\
 \text{others}=0,
\end{equation}
together with the Hamiltonian
\begin{equation}
 \hat{H}_{\mathrm{FP}}=\int\!\!d^{3}k\,\vert\bbox{k}\vert\left\{
 i(\hat{b}^{\dagger}(\bbox{k})\hat{d}(\bbox{k})-
 \hat{d}^{\dagger}(\bbox{k})\hat{b}(\bbox{k}))-1\right\}.
\end{equation}

According to the first-ordered nature of the kinetic part for $A_{0}(x)$
and $B(x)$, we meet two second class constraints in obtaining the
Hamiltonian. We treat them in the similar way as we have done for the toy
model to find
\begin{equation}
\label{eq:gaugeham0}
 \mathcal{H}_{\mathrm{G}}(x)=-\frac{1}{2}\dot{\hat{B}}(x)
 \left(\frac{1}{\nabla^{2}}\dot{\hat{B}}(x)\right)-
 \dot{\hat{B}}(x)\hat{A}_{0}(x)-\nabla \hat{B}(x)\cdot\nabla \hat{A}(x)-
 \frac{\alpha}{2}\hat{B}^{2}(x)
\end{equation}
from $\mathcal{L}_{\mathrm{G}}$. Here a comment will be in order. In
Eq.~\eqref{eq:gaugeham0}, there appears the inverse of the Laplacian that
is usually defined by
\begin{equation}
 -\frac{1}{\nabla^{2}}(\bbox{x},\bbox{y})=
 \frac{1}{4\pi}\frac{1}{\vert\bbox{x}-\bbox{y}\vert}
\end{equation}
hence causes non-locality and behaves singular at $\bbox{x}=\bbox{y}$
but we must remember that our initial Lagrangian $\mathcal{L}_{0}{'}$
does not includes the zero-modes of $A(x)$ and $A_{0}(x)$ from very
definition. Hence we should understand it by a regularized one, with
an infrared cut off parameter $\epsilon$, given explicitly in Fourier
expansion by
\begin{equation}
 -\frac{1}{\nabla^{2}}(\bbox{x})=\lim\limits_{\epsilon\to+0}
 -\frac{1}{\nabla^{2}_{\epsilon}}(\bbox{x})=
 \int_{\epsilon}\!\!\frac{d^{3}k}{(2\pi)^{3}}\,
 \frac{e^{i\bbox{k}\cdot\bbox{x}}}{\bbox{k}^{2}},
\end{equation}
in which $\epsilon$ as the lower limit of integration designates the
condition $\vert\bbox{k}\vert\ge\epsilon$. We may also need this cut off
prescription when defining creation and annihilation operators for
fields under consideration in the following. As was already stated in
section~\ref{sec:toymodel}, we also need the same regularization for
physical variables because zero-modes of massless fields cannot be
quantized in a Fock space. Hence above prescription\cite{NN} for
infrared problem applies to both physical and unphysical degrees of
the gauge field. Keeping these in mind, we consider Hamiltonian
\begin{equation}
 \hat{H}_{\mathrm{G}}=\int\!\!d^{3}x\,\mathcal{H}_{\mathrm{G}}(x)
\end{equation}
with the following equal-time commutators
\begin{equation}
 [\hat{A}(x_{0},\bbox{x}),-\dot{\hat{B}}(x_{0},\bbox{y})]=
 [\hat{A}_{0}(x_{0},\bbox{x}),\hat{B}(x_{0},\bbox{y})]=
 i\delta^{3}(\bbox{x}-\bbox{y}),\
 \text{others}=0.
\end{equation}
Heisenberg equations obtained from the Hamiltonian $\hat{H}_{\mathrm{G}}$ is identical
to the Euler-Lagrange equation and reduces, by Fourier transform, to the
one we have studied in subsection~\ref{ssec:ccr4gauge}. Hence we put
\begin{equation}
\begin{aligned}
 \hat{A}(x;\epsilon)=&\int\frac{d^{3}p}{\sqrt{(2\pi)^{3}2\vert\bbox{p}\vert}}\,
 \left\{\hat{\mathrm{a}}_{\bbox{p}}(x_{0};\epsilon)e^{i\bbox{p}\cdot\bbox{x}}+
 \hat{\mathrm{a}}^{\dagger}_{\bbox{p}}(x_{0};\epsilon)
 e^{-i\bbox{p}\cdot\bbox{x}}\right\},\\
 \hat{A}_{0}(x;\epsilon)=&\int\frac{d^{3}p}
 {\sqrt{(2\pi)^{3}2\vert\bbox{p}\vert}}\,
 \left\{\hat{\mathrm{a}}_{0\bbox{p}}(x_{0};\epsilon)e^{i\bbox{p}\cdot\bbox{x}}+
 \hat{\mathrm{a}}^{\dagger}_{0\bbox{p}}(x_{0};\epsilon)
 e^{-i\bbox{p}\cdot\bbox{x}}\right\},\\
 \hat{B}(x;\epsilon)=&\int\frac{d^{3}p}{\sqrt{(2\pi)^{3}2\vert\bbox{p}\vert}}\,
 \left\{\hat{\mathrm{b}}_{\bbox{p}}(x_{0};\epsilon)e^{i\bbox{p}\cdot\bbox{x}}+
 \hat{\mathrm{b}}^{\dagger}_{\bbox{p}}(x_{0};\epsilon)
 e^{-i\bbox{p}\cdot\bbox{x}}\right\},
\end{aligned}
\end{equation}
to find the following equal-time commutation relations of creation and
annihilation operators
\begin{equation}
\begin{aligned}
{} [\hat{\mathrm{a}}_{\bbox{p}}(x_{0};\epsilon),
 \hat{\mathrm{a}}^{\dagger}_{\bbox{q}}(x_{0};\epsilon)]=&
 \frac{1+\alpha}{2\vert\bbox{p}\vert^{2}}
 \delta_{\epsilon}^{3}(\bbox{p}-\bbox{q}),\\
 [\hat{\mathrm{a}}_{0\bbox{p}}(x_{0};\epsilon),
 \hat{\mathrm{a}}^{\dagger}_{0\bbox{q}}(x_{0};\epsilon)]=&
 -\frac{1+\alpha}{2}
 \delta_{\epsilon}^{3}(\bbox{p}-\bbox{q}),\\
 [\hat{\mathrm{a}}_{\bbox{p}}(x_{0};\epsilon),
 \hat{\mathrm{b}}^{\dagger}_{\bbox{q}}(x_{0};\epsilon)]=&
 -\delta_{\epsilon}^{3}(\bbox{p}-\bbox{q}),\\
 [\hat{\mathrm{a}}_{0\bbox{p}}(x_{0};\epsilon),
 \hat{\mathrm{b}}^{\dagger}_{\bbox{q}}(x_{0};\epsilon)]=&
 i\vert\bbox{p}\vert\delta_{\epsilon}^{3}(\bbox{p}-\bbox{q}),\\
 [\hat{\mathrm{a}}_{\bbox{p}}(x_{0};\epsilon),
 \hat{\mathrm{a}}^{\dagger}_{0\bbox{q}}(x_{0};\epsilon)]=&0,
 \end{aligned}
\end{equation}
where $\delta_{\epsilon}^{3}(\bbox{p}-\bbox{q})$ is a delta function
accompanied with a step function to achieve the above mentioned
regularization:\cite{NN} 
\begin{equation}
 \delta_{\epsilon}^{3}(\bbox{p}-\bbox{q})\equiv
 \theta(\vert\bbox{p}\vert-\epsilon)\delta^{3}(\bbox{p}-\bbox{q}).
\end{equation}
We then define following operators and their hermitian conjugates
\begin{equation}
\label{eq:gaugeoprtr}
 \hat{\mathcal{B}}_{\epsilon}(x_{0},\bbox{p})=
 \frac{1}{\vert\bbox{p}\vert}\hat{\mathrm{b}}_{\bbox{p}}(x_{0};\epsilon),\
 \hat{\mathcal{D}}_{\epsilon}(x_{0},\bbox{p})=
 \frac{1}{2}\left\{i\vert\bbox{p}\vert\hat{\mathrm{a}}_{\bbox{p}(x_{0};\epsilon)}-
 \hat{\mathrm{a}}_{0\bbox{p}}(x_{0};\epsilon)\right\}
\end{equation}
together with
\begin{equation}
\label{eq:gaugesol}
 \hat{\bar{\mathcal{D}}}_{\epsilon}(x_{0},\bbox{p})=
 \frac{1}{2}\left\{i\vert\bbox{p}\vert
 \hat{\mathrm{a}}_{\bbox{p}}(x_{0};\epsilon)+
 \hat{\mathrm{a}}_{0\bbox{p}}(x_{0};\epsilon)\right\}=
 -\frac{i}{4}(1+\alpha)\hat{\mathcal{B}}_{\epsilon}(x_{0},\bbox{p})
\end{equation}
as the analogue of Eq.~\eqref{eq:galgop} and Eq.~\eqref{eq:soldbar}.

Inversion of Eq.~\eqref{eq:gaugeoprtr} for finite $\epsilon$ with the aid
of Eq.~\eqref{eq:gaugesol} bring us
\begin{equation}
\begin{aligned}
 \hat{\mathrm{a}}_{\bbox{p}}(x_{0};\epsilon)=&
 -\frac{1}{\vert\bbox{p}\vert}\left(
 \frac{1+\alpha}{4}\hat{\mathcal{B}}_{\epsilon}(x_{0},\bbox{p})+
 i\hat{\mathcal{D}}_{\epsilon}(x_{0},\bbox{p})\right),\\
 \hat{\mathrm{a}}_{0\bbox{p}}(x_{0};\epsilon)=&
 -i\left(\frac{1+\alpha}{4}\hat{\mathcal{B}}_{\epsilon}(x_{0},\bbox{p})-
 i\hat{\mathcal{D}}_{\epsilon}(x_{0},\bbox{p})\right),\\
 \hat{\mathrm{b}}_{\bbox{p}}(x_{0};\epsilon)=&
 \vert\bbox{p}\vert\hat{\mathcal{B}}_{\epsilon}(x_{0},\bbox{p}).
\end{aligned}
\end{equation}
Then we are able to rewrite $\hat{A}(x;\epsilon)$,
$\hat{A}_{0}(x;\epsilon)$ and 
$\hat{B}(x;\epsilon)$ in terms of these creation and annihilation
operators. We thus obtain, by putting $\epsilon\to+0$ after all
calculation, that the Hamiltonian $\hat{H}_{\mathrm{G}}$ can be expressed as
\begin{equation}
\begin{aligned}
 \hat{H}_{\mathrm{G}}=&\int\!\!d^{3}p\,\vert\bbox{p}\vert
 \left\{\vphantom{\frac{1-\alpha}{2}}i\left(
 \hat{\mathcal{B}}^{\dagger}(x_{0},\bbox{p})\hat{\mathcal{D}}(x_{0},\bbox{p})-
 \hat{\mathcal{D}}^{\dagger}(x_{0},\bbox{p})\hat{\mathcal{B}}(x_{0},\bbox{p})
 \right)+1\right.\\
 &\hphantom{\int\!\!d^{3}p\,}\left.
 +\frac{1-\alpha}{2}\hat{\mathcal{B}}^{\dagger}(x_{0},\bbox{p})
 \hat{\mathcal{B}}(x_{0},\bbox{p})
 \right\}.
\end{aligned}
\end{equation}
Combining this together with that of FP ghosts, we finally obtain the
Hamiltonian for whole system, given by
\begin{equation}
\begin{aligned}
 \hat{H}=&\int\!\!d^{3}p\,\vert\bbox{p}\vert
 \left\{\vphantom{\frac{1-\alpha}{2}}i\left(
 \hat{\mathcal{B}}^{\dagger}(\bbox{p})\hat{\mathcal{D}}(\bbox{p})-
 \hat{\mathcal{D}}^{\dagger}(\bbox{p})\hat{\mathcal{B}}(\bbox{p})\right)+
 i(\hat{b}^{\dagger}(\bbox{p})\hat{d}(\bbox{p})-
 \hat{d}^{\dagger}(\bbox{p})\hat{b}(\bbox{p}))\right.\\
 &\hphantom{\int\!\!d^{3}p\,}\left.
 +\frac{1-\alpha}{2}\hat{\mathcal{B}}^{\dagger}(\bbox{p})
 \hat{\mathcal{B}}(\bbox{p})
 \right\}
\end{aligned}
\end{equation}
in which we have made use of operators in Schr\"odinger picture.
The observation here is that we can regard the system of unphysical
degrees of the gauge field with FP ghosts under a covariant gauge
condition as a collection of infinitely many copies of the system we
studied in the preceding sections through the toy model. Hence we
already know the structure, including its metric, of the Fock space
equipped with this system and how to define coherent states for
constructing a path integral in terms of them. We may, therefore, list
here main results obtained in this study:
\begin{enumerate}
 \item The Fock space is spanned by state vectors given by
\begin{equation}
 \prod_{\bbox{k}\ne0}\vert[m_{2}m_{1}]\{n_{2}n_{1}\};{\bbox{k}}\rangle,\
 m_{i}=0,\,1,\,2,\,\dots,\
 n_{i}=0,\,1\
 \text{for each }\bbox{k}.
\end{equation}

 \item The conjugation operator defined in Eq.~\eqref{eq:conjugate} is
generalized to
\begin{equation}
\label{eq:conjugateA}
\begin{aligned}
 \hat{\eta}=&\prod_{\bbox{k}\ne0}
 \exp\left[\frac{i\pi}{2}\left\{
 (\hat{\mathcal{B}}^{\dagger}(\bbox{k})+i\hat{\mathcal{D}}^{\dagger}(\bbox{k}))
 (\hat{\mathcal{B}}(\bbox{k})-i\hat{\mathcal{D}}(\bbox{k}))\right\}\right]\\
 &\times
 \exp\left[\left\{\frac{i\pi}{2}
 (\hat{b}^{\dagger}(\bbox{k})+i\hat{d}^{\dagger}(\bbox{k}))
 (\hat{b}(\bbox{k})-i\hat{d}(\bbox{k}))\right\}\right],
\end{aligned}
\end{equation}
which brings above basic vectors to their conjugates with respect to the
       inner product.

 \item A coherent state, that is a simultaneous eigenvector of
       annihilation operators $\hat{\mathcal{B}}(\bbox{k})$,
       $\hat{\mathcal{D}}(\bbox{k})$, $\hat{b}(\bbox{k})$ and
       $\hat{d}(\bbox{k})$ for all $\bbox{k}$, can be defined by
\begin{equation}
\begin{aligned} 
 \vert\{\bbox{z},\bbox{\xi}\}\rangle\equiv&
 \prod_{\bbox{k}\ne0}
 \exp\left[i\left\{
 (\hat{\mathcal{B}}^{\dagger}(\bbox{k})z_{1}(\bbox{k})+
 \hat{\mathcal{D}}^{\dagger}(\bbox{k})z_{2}(\bbox{k}))\right\}\right]\\
 &\times
 \exp\left[i\left\{
 (\hat{b}^{\dagger}(\bbox{k})\xi_{1}(\bbox{k})+
 \hat{d}^{\dagger}(\bbox{k})\xi_{2}(\bbox{k}))\right\}\right]
 \vert0\rangle.
\end{aligned}
\end{equation}

 \item Left eigenvectors of creation operators are obtained by usual
       hermitian conjugation followed by a multiplication of the
       conjugation operator: 
\begin{equation}
 \langle\underline{\{\bbox{z},\bbox{\xi}\}}\vert=
 (\vert\{\bbox{z},\bbox{\xi}\}\rangle)^{\dagger}\hat{\eta}
\end{equation}
to yield an inner product
\begin{equation}
 \langle\underline{\{\bbox{z},\bbox{\xi}\}}\vert
 \{\bbox{z}{'},\bbox{\xi}{'}\}\rangle=
 \exp\left[\int\!\!d^{3}k\,\left\{
 \bbox{z}^{\dagger}(\bbox{k})\bbox{z}{'}(\bbox{k})+
 \bbox{\xi}^{\dagger}(\bbox{k})\bbox{\xi}{'}(\bbox{k})\right\}\right].
\end{equation}

 \item The set of coherent states gives a resolution of unity
\begin{equation}
\begin{aligned}
 &\int\prod_{\bbox{k}\ne0}
 \left[\left(\frac{dz(\bbox{k})\,dz^{*}(\bbox{k})}{\pi}\right)^{2}
 (d\xi(\bbox{k})\,d\xi^{*}(\bbox{k}))^{2}\right]\,
\\
 &\times
 \exp\left[-\int\!\!d^{3}k\,\left\{
 \bbox{z}^{\dagger}(\bbox{k})\bbox{z}(\bbox{k})+
 \bbox{\xi}^{\dagger}(\bbox{k})\bbox{\xi}(\bbox{k})\right\}\right]
 \vert\{\bbox{z},\bbox{\xi}\}\rangle
 \langle\{\bbox{z},\bbox{\xi}\}\vert=1.
\end{aligned}
\end{equation}

 \item Only the vacuum specified by
\begin{equation}
 \vert0\rangle\equiv\prod_{\bbox{k}\ne0}\vert[00]\{00\};{\bbox{k}}\rangle
\end{equation}
can be a positive normed physical state, i.e. BRS singlet.

 \item The normalized BRS charge is given by
\begin{equation}
 \hat{Q}_{B}=\int\!\!d^{3}k\,i\left\{
 \hat{\mathcal{B}}(\bbox{k})\hat{b}^{\dagger}(\bbox{k})-
 \hat{\mathcal{B}}^{\dagger}(\bbox{k})\hat{b}(\bbox{k})\right\}
\end{equation}
and accompanied with another operator, that is conserved only in the
       Feynman gauge, $\hat{Q}_{D}=\hat{\eta}\hat{Q}_{B}\hat{\eta}$ that
       generates the BRS-inversion and is given by
\begin{equation}
 \hat{Q}_{D}=\int\!\!d^{3}k\,i\left\{
 \hat{\mathcal{D}}(\bbox{k})\hat{d}^{\dagger}(\bbox{k})-
 \hat{\mathcal{D}}^{\dagger}(\bbox{k})\hat{d}(\bbox{k})\right\}.
\end{equation}

 \item BRS variant member of a BRS quartet is a partner state of BRS exact
       member and is BRS-inversion exact. In other word, BRS daughter
       state is BRS-inversion parent state.

 \item An operator that counts total number of excitations in all BRS
       quartet is found to be
\begin{equation}
 \hat{N}=\int\!\!d^{3}p\,\left\{i\left(
 \hat{\mathcal{B}}^{\dagger}(\bbox{p})\hat{\mathcal{D}}(\bbox{p})-
 \hat{\mathcal{D}}^{\dagger}(\bbox{p})\hat{\mathcal{B}}(\bbox{p})\right)+
 i(\hat{b}^{\dagger}(\bbox{p})\hat{d}(\bbox{p})-
 \hat{d}^{\dagger}(\bbox{p})\hat{b}(\bbox{p}))\right\}.
\end{equation}
Another expression for this operator is given by
\begin{equation}
 \hat{N}=\{\hat{Q}_{B},\hat{Q}_{D}\}.
\end{equation}

 \item Kugo-Ojima projection is constructed explicitly as
\begin{equation}
 P^{(n)}=\frac{1}{2\pi}\int_{0}^{2\pi}\!\!d\theta\,
 e^{i\theta(\hat{N}-n)},\
 n=0,\,1,\,2,\,\dots\,.
\end{equation}
It can be expressed in terms of coherent states as
\begin{equation}
\begin{aligned}
 P^{(n)}=&\frac{1}{2\pi}\int_{0}^{2\pi}\!\!d\theta\,
 e^{-in\theta}
 \int\prod_{\bbox{k}\ne0}
 \left[\left(\frac{dz(\bbox{k})\,dz^{*}(\bbox{k})}{\pi}\right)^{2}
 (d\xi(\bbox{k})\,d\xi^{*}(\bbox{k}))^{2}\right]\,
\\
 &\times
 \exp\left[-\int\!\!d^{3}k\,\left\{
 \bbox{z}^{\dagger}(\bbox{k})\bbox{z}(\bbox{k})+
 \bbox{\xi}^{\dagger}(\bbox{k})\bbox{\xi}(\bbox{k})\right\}\right]
 \vert\{e^{i\theta}\bbox{z},e^{i\theta}\bbox{\xi}\}\rangle
 \langle\underline{\{\bbox{z},\bbox{\xi}\}}\vert.
\end{aligned}
\end{equation}

\end{enumerate}

Besides these fundamental properties, we should confirm the usefulness and
reliability of our method of constructing a path integral by means of
coherent states. As will be expected from the facts we have seen in this
section, generating functionals, Eq.~\eqref{eq:genfunc01} for FP ghost and
Eq.~\eqref{eq:genfunc02} for unphysical degrees of the gauge field, can be
immediately generalized to the current case: generating functionals for
the system under consideration is defined by inserting sources
$\bar{\eta}(x)$, $\eta(x)$, $J(x)$, $J_{0}(x)$ and $J_{B}(x)$ against
$\hat{c}(x)$, $\hat{\bar{c}}(x)$, $\hat{A}(x)$, $\hat{A}_{0}(x)$ and
$\hat{B}(x)$ in the same manner as has been done in
section~\ref{sec:gfefa} to be evaluated as
\begin{equation}
\label{eq:genfunc03}
 W^{(\mathrm{FP})}[\eta,\eta^{\dagger}]=
 -\frac{1}{2}\int\!\!d^{4}x\,d^{4}x{'}\,
 \frac{1}{i}\Delta_{F}(x-x{'})
 \left\{\bar{\eta}(x)\eta(x{'})-\eta(x)\bar{\eta}(x{'})\right\},
\end{equation}
\begin{equation}
\begin{gathered}
\label{eq:genfunc04}
 W^{(\mathrm{G})}[J,J_{0},J_{B}]=-\frac{1}{2}\int\!\!d^{4}x\,d^{4}x{'}\,
 \bbox{J}^{T}(x)D^{(\mathrm{G})}_{F}(x-x{'})\bbox{J}(x{'}),\\
 \bbox{J}^{T}(x)=(J(x),J_{0}(x),J_{B}(x)),
\end{gathered}
\end{equation}
in which propagators, $-i\Delta_{F}(x)$ for FP ghosts and
$D^{(\mathrm{G})}_{F}(x)$ for the gauge field, are defined by
\begin{equation}
\begin{gathered}
 -i\Delta_{F}(x)=\int\!\!\frac{d^{4}p}{(2\pi)^{4}}\,
 e^{-ipx}\frac{-i}{p^{2}},\\
 D^{(\mathrm{G})}_{F}(x)=\int\!\!\frac{d^{4}p}{(2\pi)^{4}}\,
 e^{-ipx}\tilde{D}^{(\mathrm{G})}_{F}(p),\\
 \tilde{D}^{(\mathrm{G})}_{F}(p)=\frac{1}{p^{2}}
 \begin{pmatrix}
  \dfrac{1}{\bbox{p}^{2}}\left\{1-(1-\alpha)
  \dfrac{\bbox{p}^{2}}{p^{2}}\right\}&
  (1-\alpha)\dfrac{p_{0}}{p^{2}}&-1\\
  -(1-\alpha)\dfrac{p_{0}}{p^{2}}&
  -1+(1-\alpha)\dfrac{p_{0}^{2}}{p^{2}}&-p_{0}\\
  -1&p_{0}&0
 \end{pmatrix},
\end{gathered}
\end{equation}
where $p^{2}=p_{0}^{2}+\bbox{p}^{2}$. The generating functional for
FP ghosts given by Eq.~\eqref{eq:genfunc03} will be immediately brought to its
corresponding effective action
\begin{equation}
 \varGamma_{E}^{(\mathrm{FP})}=\int\!\!d^{4}x\,
 i\partial_{\mu}\bar{c}(x)\partial^{\mu}c(x)
 \xrightarrow{\ x_{0}\to ix_{0}\ }
 i\varGamma^{(\mathrm{FP})}=i\int\!\!d^{4}x\,
 \left\{-i\partial_{\mu}\bar{c}(x)\partial^{\mu}c(x)\right\}
\end{equation}
to yield the classical action of FP ghosts in the Minkowski
space-time. We must next combine generating functional for unphysical
part of the gauge field with the one from physical degrees. To this aim
we rearrange external sources for $\hat{A}^{\mu}(x)$ to be
$j_{\mu}(x)$. Since $J_{0}(x)\hat{A}^{0}(x)$ is already included, we
just set $j_{0}(x)=J_{0}(x)$ and concentrate on the spatial components
$-\bbox{j}(x)\cdot\hat{\bbox{A}}(x)$. From our definition, this term is
rewritten as
\begin{equation}
 -\bbox{j}(x)\cdot\hat{\bbox{A}}(x)=-(\nabla\cdot\bbox{j}(x))\hat{A}(x)-
 \bbox{j}(x)\cdot\hat{\bbox{A}}_{\mathrm{T}}(x).
\end{equation}
Hence we put $J(x)=-\nabla\cdot\bbox{j}(x)$ in Eq.~\eqref{eq:genfunc04}
and combine it with the contribution from physical degrees, given by
\begin{equation}
\label{eq:genfunc05}
 W^{(\mathrm{T})}[\bbox{j}]=-\frac{1}{2}\int\!\!d^{4}x\,d^{4}x{'}\,
 \bbox{j}^{T}(x)D^{(\mathrm{T})}_{F}(x-x{'})\bbox{j}(x{'}),
\end{equation}
where $D^{(\mathrm{T})}_{F}(x-x{'})$ is defined by
\begin{equation}
 D^{(\mathrm{T})}_{F}(x-x{'})=\int\!\!\frac{d^{4}p}{(2\pi)^{4}}\,
 e^{-ipx}\frac{1}{p^{2}}
 \left(\mathbf{1}-\frac{\bbox{p}\bbox{p}^{T}}{\bbox{p}^{2}}\right).
\end{equation}
Thus total generating functional for the gauge field is obtained as
\begin{equation}
\label{eq:genfunc06}
 \begin{aligned}
 W[j_{\mu},J_{B}]=
 &W^{(\mathrm{G})}[j_{\mu},J_{B}]+W^{(\mathrm{T})}[\bbox{j}]\\
 =&
 -\frac{1}{2}\int\!\!d^{4}x\,d^{4}x{'}\,
 \bbox{J}^{T}(x)D^{(\alpha)}_{F}(x-x{'})\bbox{J}(x{'}),
 \end{aligned}
\end{equation}
in which $\bbox{J}^{T}(x)=(\bbox{j}^{T}(x),j_{0}(x),J_{B}(x))$ and the
propagator is given by
\begin{equation}
\label{eq:propagator}
\begin{gathered}
 D^{(\alpha)}_{F}(x)=\int\!\!\frac{d^{4}p}{(2\pi)^{4}}\,
 e^{-ipx}\tilde{D}^{(\alpha)}_{F}(p),\\
 \tilde{D}^{(\alpha)}_{F}(p)=\frac{1}{p^{2}}
 \begin{pmatrix}
  \mathbf{1}-(1-\alpha)\dfrac{\bbox{p}\bbox{p}^{T}}{p^{2}}&
  -i(1-\alpha)\dfrac{p_{0}\bbox{p}}{p^{2}}&i\bbox{p}\\
  -i(1-\alpha)\dfrac{p_{0}\bbox{p}^{T}}{p^{2}}&
  -1+(1-\alpha)\dfrac{p_{0}^{2}}{p^{2}}&-p_{0}\\
  -i\bbox{p}^{T}&p_{0}&0
 \end{pmatrix}.
\end{gathered}
\end{equation}
If we perform the inverse Wick rotation,
$x_{0}\to ix_{0}$, $x_{0}{'}\to ix_{0}{'}$ 
in Eq.~\eqref{eq:genfunc06} in addition to $p_{0}\to-ip_{0}$
in the propagator in Eq.~\eqref{eq:propagator}, we can immediately find
that the Euclidean generating functional given by
Eq.~\eqref{eq:genfunc06} yields the corresponding one for the abelian
gauge field with a covariant gauge condition in Minkowski space-time. The
Euclidean effective action for the gauge field is also found from
Eq.~\eqref{eq:genfunc06} to be
\begin{equation}
\begin{aligned}
 \varGamma_{E}=&\int\!\!d^{4}x\,\frac{1}{2}\left\{
 (i\dot{\bbox{A}}(x)+\nabla A_{0}(x))^{2}-(\nabla\times\bbox{A}(x))\right.\\
 &\left.
 -2i\dot{B}(x)A_{0}+2B(x)\nabla\cdot\bbox{A}(x)+\alpha B^{2}(x)\right\}
\end{aligned}
\end{equation}
which can be translated into Minkowski effective action by inverse Wick
rotation: 
\begin{equation}
\begin{aligned}
 \varGamma_{E}\xrightarrow{\ x_{0}\to ix_{0}\ }
 i\varGamma=
 &i\int\!\!d^{4}x\,\frac{1}{2}\left\{
 (\dot{\bbox{A}}(x)+\nabla A_{0}(x))^{2}-(\nabla\times\bbox{A}(x))\right.\\
 &\left.
 -2\dot{B}(x)A_{0}+2B(x)\nabla\cdot\bbox{A}(x)+\alpha B^{2}(x)\right\}.
\end{aligned}
\end{equation}
Hence restoration of Lorentz covariance is evident.
We thus confirmed that our prescription for constructing a path
integral of a free gauge field from manifestly covariant operator
formalism in terms of coherent states works quite fine to provide
essential ingredients for perturbative expansion in its zeroth order in
entirely covariant manner.

\section{Field diagonal representation}
\label{sec:discussion}
Our considerations has been so far restricted to the construction and
its use of the coherent state for unphysical degrees of a gauge field in
a unified manner with those for FP ghosts. As for ghost fermions, there
is a path integral in terms of field eigenvectors.\cite{TKfermi} \ 
It will be, therefore, beautiful if we can formulate a path integral in
terms of eigenvectors of field operators and their canonical
conjugates in entirely covariant way. To discuss such problem as a whole
is, however, beyond the scope of this paper. We here consider a field
diagonal representation only for the Feynman gauge to fill the
discrepancy of the results in Ref.~\citen{KS} \ and Ref.~\citen{AFIO}.

In the covariant path integral for the gauge field of Ref.~\citen{KS}, \ 
$A_{0}$($A_{4}$) is treated as an auxiliary field that disappears once
from the formulation but comes back into the path integral by use of the
Gaussian identity. The technique utilized there is the one for
functional analysis on a Hilbert space with positive definite metric.
On the other hand, the basic ingredient of the construction in
Ref.~\citen{AFIO} \ is the use of eigenvector of $\hat{A}_{0}$ defined
on a representation space with indefinite metric. So let us reconsider
the prescription of Ref.~\citen{AFIO}\  from the viewpoint of our
standpoint in this paper.

Returning back to the toy model again and putting here $\alpha=1$ to
restrict ourselves to the Feynman gauge, let us consider an operator
\begin{equation}
 C\int\!\!d\mu\,d\nu\,
 e^{\mu (\hat{A}-q)+\nu (\hat{A}_{0}-q_{0})}
\end{equation}
in which ranges of integrations with respect to $\mu$ and $\nu$ together
with the normalization factor $C$ should
be determined so that integrations with respect to $q$ and $q_{0}$ of
above introduced operator yield an expression of the resolution of unity.
Note that the construction of such an expression in terms of
Schr\"odinger operators will be immediately translated to that of
Heisenberg operators. Making use of relations $\hat{A}$, $\hat{A}_{0}$
and $\hat{B}$ with the creation and annihilation operators
$\hat{\mathcal{B}}$, $\hat{\mathcal{D}}$, $\hat{\mathcal{B}}^{\dagger}$
and $\hat{\mathcal{D}}^{\dagger}$, we can easily obtain
\begin{equation}
\label{eq:resuni01}
\begin{aligned}
 &\int_{-\infty}^{\infty}\!\!\frac{d\mu\,d\nu}{(2\pi)^{2}}\,
 \langle\underline{\bbox{z}}\vert
 e^{i\mu (\hat{A}-q)+\nu (\hat{A}_{0}-iq_{0})}\vert\bbox{z}{'}\rangle\\
 =&
 \frac{k^{2}}{\pi}
 \exp\left[-k^{3}\left\{
 q-\frac{i}{\sqrt{2k^{3}}}\left(
 \frac{1}{2}(z_{2}^{*}+iz_{1}{'})-i(z_{1}^{*}-iz_{2}{'})
 \right)\right\}^{2}\right]\\
 &\times
 \exp\left[-k\left\{
 q_{0}+\frac{i}{\sqrt{2k}}\left(
 \frac{1}{2}(z_{2}^{*}-iz_{1}{'})+i(z_{1}^{*}+iz_{2}{'})
 \right)\right\}^{2}\right]
 e^{\bbox{z}^{\dagger}\bbox{z}{'}}
\end{aligned}
\end{equation}
for an arbitrary pair of coherent states. From this calculation, we read
\begin{equation}
 \int_{-\infty}^{\infty}\!\!dq\,dq_{0}\,
 \int_{-\infty}^{\infty}\!\!\frac{d\mu\,d\nu}{(2\pi)^{2}}\,
 \langle\underline{\bbox{z}}\vert
 e^{i\mu (\hat{A}-q)+\nu (\hat{A}_{0}-iq_{0})}\vert\bbox{z}{'}\rangle=
 \langle\underline{\bbox{z}}\vert\bbox{z}{'}\rangle
\end{equation}
which implies
\begin{gather}
 \int_{-\infty}^{\infty}\!\!dq\,dq_{0}\,
 \vert{q,q_{0}}\rangle\langle\underline{q,q_{0}}\vert=1,\\
 \vert{q,q_{0}}\rangle\langle\underline{q,q_{0}}\vert=
 \int_{-\infty}^{\infty}\!\!\frac{d\mu\,d\nu}{(2\pi)^{2}}\,
 e^{i\mu (\hat{A}-q)+\nu (\hat{A}_{0}-iq_{0})}.
\label{eq:qidentity}
\end{gather}
To find the explicit form of the eigenvector $\vert{q,q_{0}}\rangle$ and
its conjugate, we calculate the projection
$\vert{q,q_{0}}\rangle\langle\underline{q,q_{0}}\vert$ defined above by
\begin{equation}
 \int\!\!\left(\frac{dz\,dz^{*}}{\pi}\right)^{2}\,
 \left(\frac{dz{'}\,dz^{*}{'}}{\pi}\right)^{2}\,
 e^{-\bbox{z}^{\dagger}\bbox{z}-\bbox{z}^{\dagger}{'}\bbox{z}{'}}
 \vert\bbox{z}\rangle\langle\underline{\bbox{z}}\vert{q,q_{0}}\rangle
 \langle\underline{q,q_{0}}\vert\bbox{z}{'}\rangle
 \langle\underline{\bbox{z}{'}}\vert.
\end{equation}
It is straightforward to obtain
\begin{equation}
\begin{aligned}
 \vert{q,q_{0}}\rangle=&
 \int\!\!\left(\frac{dz\,dz^{*}}{\pi}\right)^{2}\,
 e^{-\bbox{z}^{\dagger}\bbox{z}}
 \vert\bbox{z}\rangle\langle\underline{\bbox{z}}\vert{q,q_{0}}\rangle\\
 =&
 \frac{k}{\sqrt{\pi}}
 \exp\left\{-\frac{k}{2}(k^{2}q^{2}+q_{0}^{2})\right\}
 \\
 &\times
 \exp\left[-\sqrt{\frac{k}{2}}(kq-q_{0})\hat{\mathcal{B}}^{\dagger}+
 i\sqrt{2k}(kq+q_{0})\hat{\mathcal{D}}^{\dagger}
 -\frac{1}{4}\hat{\mathcal{B}}^{\dagger}{}^{2}+
 \hat{\mathcal{D}}^{\dagger}{}^{2}\right]\vert0\rangle
\end{aligned}
\end{equation}
and its conjugate
\begin{equation}
\begin{aligned}
 \langle\underline{q,q_{0}}\vert=&
 \frac{k}{\sqrt{\pi}}
 \exp\left\{-\frac{k}{2}(k^{2}q^{2}+q_{0}^{2})\right\}
 \\
 &\times
 \langle0\vert
 \exp\left[-i\sqrt{\frac{k}{2}}(kq-q_{0})\hat{\mathcal{D}}
 -\sqrt{2k}(kq+q_{0})\hat{\mathcal{B}}
 +\frac{1}{4}\hat{\mathcal{D}}^{2}-
 \hat{\mathcal{B}}^{2}\right].
\end{aligned}
\end{equation}
Thus we have found a simultaneous eigenvectors of $\hat{A}$ and
$\hat{A}_{0}$ because they satisfy
\begin{equation}
\begin{gathered}
 \hat{A}\vert{q,q_{0}}\rangle=q\vert{q,q_{0}}\rangle,\
 \hat{A}_{0}\vert{q,q_{0}}\rangle=iq_{0}\vert{q,q_{0}}\rangle,\\
 \langle\underline{q,q_{0}}\vert \hat{A}=q\langle\underline{q,q_{0}}\vert,\
 \langle\underline{q,q_{0}}\vert \hat{A}_{0}=
 iq_{0}\langle\underline{q,q_{0}}\vert
\end{gathered}
\label{eq:qeig}
\end{equation}
by definition. It is easy to see that an inner product of these
eigenvectors are given by
\begin{equation}
 \langle\underline{q,q_{0}}\vert{q{'},q{'}_{0}}\rangle=
 \delta(q-q{'})\delta(q_{0}-q{'}_{0}).
\end{equation}

Turning now to the eigenvectors of canonical conjugates to $\hat{A}$ and
$\hat{A}_{0}$, we may examine
\begin{equation}
 \int\!\!d\mu\,d\nu\,e^{\mu (\hat{P}-p)+\nu (\hat{P}_{0}-p_{0})}
\end{equation}
in the same way we have done above. It will immediately fail, however,
to yield simultaneous eigenvectors of $\hat{P}$ and 
$\hat{P}_{0}$ because they are expressed solely
by $\hat{\mathcal{B}}$ and $\hat{\mathcal{B}}^{\dagger}$ hence no
Gaussian normalization, which was found for $\hat{A}$ and $\hat{A}_{0}$
as is seen in Eq.~\eqref{eq:resuni01}, will be 
available. We can overcome this difficulty by introducing
$\hat{\varPi}=\hat{P}+k^{2}\hat{A}_{0}$ and
$\hat{\varPi}_{0}=-\hat{P}_{0}-k^{2}\hat{A}$ instead of $\hat{P}$ and
$\hat{P}_{0}$ to observe
\begin{gather}
 \int_{-\infty}^{\infty}\!\!dp\,dp_{0}\,
 \vert{p,p_{0}}\rangle\langle\underline{p,p_{0}}\vert=1,\\
 \vert{p,p_{0}}\rangle\langle\underline{p,p_{0}}\vert=
 \int_{-\infty}^{\infty}\!\!\frac{d\mu\,d\nu}{(2\pi)^{2}}\,
 e^{i\mu (\hat{\varPi}-p)+\nu (\hat{\varPi}_{0}-ip_{0})},\\
 \langle\underline{p,p_{0}}\vert{p{'},p{'}_{0}}\rangle=
 \delta(p-p{'})\delta(p_{0}-p{'}_{0}),
\label{eq:pidentity}
\end{gather}
and the inner products
\begin{equation}
 \langle\underline{q,q_{0}}\vert{p,p_{0}}\rangle=
 \frac{1}{2\pi}e^{ipq+ip_{0}q_{0}},\
 \langle\underline{p,p_{0}}\vert{q,q_{0}}\rangle=
 \frac{1}{2\pi}e^{-ipq-ip_{0}q_{0}},
\end{equation}
as well. Similar to the eigenvectors of $\hat{A}$ and $\hat{A}_{0}$
above, we have 
\begin{equation}
\begin{gathered}
 \hat{\varPi}\vert{p,p_{0}}\rangle=p\vert{p,p_{0}}\rangle,\
 \hat{\varPi}_{0}\vert{p,p_{0}}\rangle=ip_{0}\vert{p,p_{0}}\rangle,\\
 \langle\underline{p,p_{0}}\vert \hat{\varPi}=
 p\langle\underline{p,p_{0}}\vert,\
 \langle\underline{p,p_{0}}\vert \hat{\varPi}_{0}=
 ip_{0}\langle\underline{p,p_{0}}\vert.
\end{gathered}
\label{eq:peig}
\end{equation}
It should be noted here that our definition of these eigenvectors is
precisely same as the one in Ref.~\citen{AFIO} \ excepting that we
accept eigenvalues of $\hat{A}_{0}$ and $\hat{\varPi}_{0}$ being pure
imaginary. This causes an artificial contradiction to the hermiticity of
these operators. But we must take the metric into account in considering
hermiticity of operators on the indefinite metric vector space. Indeed,
spectral representations of these operators satisfy
\begin{equation}
\begin{aligned}
 \hat{\eta}_{\mathrm{G}}^{\dagger}\hat{A}_{0}^{\dagger}
 \hat{\eta}_{\mathrm{G}}=
 \hat{\eta}_{\mathrm{G}}^{\dagger}\left(
 \int_{-\infty}^{\infty}\!\!dq\,dq_{0}\,
 iq_{0} \vert{q,q_{0}}\rangle\langle\underline{q,q_{0}}\vert
 \right)^{\dagger}\hat{\eta}_{\mathrm{G}}=
 \hat{A}_{0},\\
 \hat{\eta}_{\mathrm{G}}^{\dagger}\hat{\varPi}_{0}^{\dagger}
 \hat{\eta}_{\mathrm{G}}=
 \hat{\eta}_{\mathrm{G}}^{\dagger}\left(
 \int_{-\infty}^{\infty}\!\!dp\,dp_{0}\,
 ip_{0} \vert{p,p_{0}}\rangle\langle\underline{p,p_{0}}\vert
 \right)^{\dagger}\hat{\eta}_{\mathrm{G}}=
 \hat{\varPi}_{0}.
\end{aligned}
\end{equation}

If we put $\alpha=1$ for the Feynman gauge in Eq.~\eqref{eq:gham}, the
Hamiltonian $\hat{H}_{\mathrm{G}}$ becomes
\begin{equation}
 \hat{H}_{\mathrm{G}}^{(F)}=\frac{1}{2k^{2}}\hat{\varPi}^{2}+
 \frac{k^{4}}{2}\hat{A}^{2}-
 \frac{1}{2}\hat{\varPi}_{0}^{2}-\frac{k^{2}}{2}\hat{A}_{0}^{2}
\label{eq:gham1}
\end{equation}
to yield a path integral
\begin{equation}
\begin{aligned}
 &\int_{-\infty}^{\infty}\!\!dq\,dq_{0}
 \langle\underline{q,q_{0};t_{F}}\vert
 \mathrm{T}\exp\left\{
 -\int_{t_{1}}^{t_{2}}\!\!dt\,
 J(t)\hat{A}(t)+J_{0}(t)\hat{A}_{0}(t)\right\}
 \vert{q,q_{0};t_{I}}\rangle\\
 =&\lim\limits_{n\to\infty}
 \int\!\!\prod_{i=1}^{n}\frac{d^{2}p(i)\,d^{2}q(i)}{(2\pi)^{2}}
 \exp\left[\sum_{k=1}^{n}\left\{\vphantom{\frac{1}{2}}
 ip(k)\Delta q(k)+
 ip_{0}(k)\Delta q_{0}(k)
 \right.\right.\\
 &\left.\left.
 -\frac{\epsilon}{2}\left(\frac{1}{k^{2}}p^{2}(k)+
 k^{4}q^{2}(k)+p_{0}^{2}+k^{2}q_{0}^{2}
 \right)
 -\epsilon(J(k)q(k)+iJ_{0}(k)q_{0}(k))
 \right\}\vphantom{\sum_{k=1}^{n}}\right],
\label{eq:qtr}
\end{aligned}
\end{equation}
where $\epsilon=(t_{F}-t_{I})/n$ and the limiting procedure,
$t_{F},-t_{I}\to\infty$ followed by $t_{2},-t_{1}\to\infty$ should be
expected. Eq.~\eqref{eq:qtr} is nothing but another path integral
representation of the generating functional for unphysical components of
the gauge field in the toy model.
On performance of the same prescription as has been done
in section~\ref{sec:gfefa}, we will obtain the same generating
functional and the effective action again. Hence it reproduces
Eq.~\eqref{eq:efacmin} through the same procedure. 

We can immediately extend the above construction to the field
theoretical situation to obtain a Euclidean path integral
\begin{equation}
 \begin{aligned}
 &\langle{0}\vert\mathrm{T}\exp\left\{
 -\int\!\!d^{4}x\,J_{\mu}(x)\hat{A}^{\mu}(x)\right\}
 \vert{0}\rangle\\
 =&\mathcal{N}\int\!\!\mathcal{D}A_{\mu}\,
 \exp\left[-\frac{1}{2}\int\!\!d^{4}x\,
 \left\{(\partial_{\mu}A_{\nu})^{2}+
 \bbox{J}(x)\cdot\bbox{A}(x)+iJ_{0}(x)A_{0}(x)\right\}
 \right],
 \end{aligned}
\label{eq:gaugegen}
\end{equation}
where $\mathcal{N}=\{\Det(-\partial_{\mu}^{2})\}^{2}$ and summations
over repeated indices are expected by assuming the Euclidean
metric. This is the generalization to the result of Arisue et al in
Ref.~\citen{AFIO} \ by the introduction of the source terms. If we put
$iJ_{0}=J_{4}$ and write $A_{0}$ as $A_{4}$, our result coincides with
$\tilde{Z}[J_{\mu};\infty]_{\mathrm{covariant}}$ given by Kashiwa and
Sakamoto in Ref.~\citen{KS}. Then we may apply the ``Euclidean
Technique'' to the above formula. There exist, however, another
prescription to deal with this generating functional; We may leave
$iJ_{0}$ as it is and carry out the same calculation as has been done in
the previous section to find
\begin{equation}
 W[J_{\mu}]=-\frac{1}{2}\int\!\!d^{4}x\,\left\{
 \bbox{J}^{T}(x)\frac{1}{-\partial_{\mu}^{2}}\bbox{J}(x)-
 J_{0}(x)\frac{1}{-\partial_{\mu}^{2}}J_{0}(x)\right\}
\end{equation}
and the effective action
\begin{equation}
 \varGamma_{E}[A_{\mu}]=\frac{1}{2}\int\!\!d^{4}x\,\left\{
 \bbox{A}^{T}(x)\partial_{\mu}^{2}\bbox{A}(x)-
 A_{0}(x)\partial_{\mu}^{2}A_{0}(x)\right\}.
\end{equation}
Then, by performing the inverse Wick rotation, we obtain the effective
action for the gauge field in Minkowski space-time:
\begin{equation}
 \varGamma[A_{\mu}]=-\frac{1}{2}\int\!\!d^{4}x\,
 (\partial_{\mu}A^{\nu}(x))(\partial^{\mu}A_{\nu}(x)),
\end{equation}
which is nothing but the classical action of the gauge field in the
Feynman gauge. In this way we see that we need only to perform Wick
rotation and its inverse to obtain a covariant result through a
well-defined path integral although we meet a non-covariant action in
the exponent of the path integral.

Let us add some remarks on the path integral given above. First, we have
not taken care about boundary conditions of the path integral excepting
the periodic one for the time axis. In order to give a precise meaning
to the above manipulation, we must define the theory in a box of finite
volume then a boundary condition must be assigned for the spatial
boundaries. In doing so we must also treat the infrared singularity in a
proper manner because we are dealing with a massless field. These issues
were beautifully resolved at once in Ref.~\citen{KS} \ by Kashiwa and
Sakamoto by means of $b$-boundary prescription. Secondly, if we consider the
Minkowski version of the Eq.~\eqref{eq:gaugegen} by inserting the vacuum
wave-functionals, we will obtain again a non-covariant action in the
exponent of the path integral because of the negative metric for
$A_{0}$. Furthermore, the source term of $\hat{A}_{0}$ contributes
$J_{0}A_{0}$ while spatial components bring to us
$i\bbox{J}\cdot\bbox{A}$ in the action of the path integral.
Since the quadratic part of the action is treated as Fresnel
integral for this case, we must regard $J_{0}$ as pure imaginary for the
convergence of the path integral. Then we may be possible to obtain the
effective action above again. It should be stressed, however,
that \emph{the Minkowski path integral thus obtained from the covariant
canonical formalism does not possess the manifest covariance}\/. 
We may replace $iA_{0}$ in the path integral by $A_{0}$ to recover the
covariant appearance. This is the meaning of the integration along an
imaginary curve by Arisue et al for $A_{0}$ in Ref.~\citen{AFIO}. \ 
The conventional path integral of gauge fields and the result of
Ref.~\citen{KS} \  for the Minkowski space-time also have manifest
covariance and integrate $A_{0}$ along real axis. To understand this
disagreement let us recall that even for a negative definite Hamiltonian
\begin{equation}
 \hat{H}=-\frac{1}{2}\hat{p}^{2}-\frac{1}{2}\hat{q}^{2},
\end{equation}
there exist a unitary representation of $e^{-it\hat{H}}$ for a real $t$
on the representation space of CCR $[\hat{q},\hat{p}]=i$, that is a
positive metric Hilbert space, though the Euclidean path integral for
$it=\beta>0$ becomes ill-defined. Manifestly covariant path integrals
for gauge fields in Minkowski space-time can be viewed as the one
obtained in this way by use of positive metric Hilbert space even for
the unphysical degrees. The Euclidean versions of these path integrals
will be then obtained on substitutions:
\begin{equation}
 ix^{0}=x^{4},\
 A_{0}=iA_{4}
\end{equation}
as path integrals of a field theory on the Euclidean space-time. It must
be remembered, however, that a path integral obtained in this way does
not have an immediate connection to the covariant canonical
formalism. This prescription will be applied to the negative oscillator
above to be seen as
\begin{equation}
 \int\!\!\mathcal{D}p\,\mathcal{D}q\,
 e^{i\int\!\!dt\,\{p\dot{q}+(p^{2}+q^{2})/2\}}=
 \int\!\!\mathcal{D}q\,
 e^{-i\int\!\!dt\,\{(\dot{q}^{2}-q^{2})/2\}}\mapsto
 \int\!\!\mathcal{D}q_{4}\,
 e^{-\int\!\!d\tau\,\{(\dot{q}_{4}^{2}+q_{4}^{2})/2\}}.
\end{equation}
On the other hand, the prescription yields the same Euclidean path
integral as the one obtained through usual ``Euclidean Technique'' for a
normal harmonic oscillator:
\begin{equation}
 \int\!\!\mathcal{D}Q\,
 e^{i\int\!\!dt\,\{(\dot{Q}^{2}-Q^{2})/2\}}\mapsto
 \int\!\!\mathcal{D}Q\,
 e^{-\int\!\!d\tau\,\{(\dot{Q}^{2}+Q^{2})/2\}}=
 \Tr(e^{-\tau(\hat{P}^{2}+\hat{Q}^{2})/2}).
\end{equation}
This clearly explains why Kashiwa and Sakamoto could obtain covariant
path integrals in the Minkowski space-time as well as the Euclidean
formulas by treating both integrations with respect to $A_{0}$ and $A_{4}$
along the real axis.

\section{Conclusion}
We have made a thorough investigation on a toy model which explains the
structure of Fock space of the quantized gauge field with a covariant
gauge condition by means of BRS formalism. A prescription for defining
coherent states both for FP ghosts and unphysical degrees of the gauge
field has been developed to achieve a construction of path integrals for
them by use of these coherent states. Coherent state path integral
constructed in this way has a concrete relation to the canonical
formulation of the theory. Hence we can always go and back between both
formulations easily. This is in sharp contrast with the situation of
conventional formulation of path integrals, that is formal functional
integrations with classical action in the exponent of its integrand, for
gauge theories. 

Although our considerations were restricted to an abelian gauge theory,
our approach will hold even for non-abelian gauge fields as far as
the zeroth order of perturbation or renormalized asymptotic fields are
concerned. Therefore we may expect, at least formally, that there
exist same structures of Fock spaces and BRS-quartets will be formulated
entirely in the same manner as for the abelian case if we express the
Lagrangian of such systems in terms of renormalized asymptotic
fields. Then it immediately follows that Kugo-Ojima projection expressed
in terms of these asymptotic fields possesses the same form as the one
constructed in this paper for the abelian case, even though practical use
of such a formula in terms of Heisenberg operators may be difficult.

As for the practical use in perturbation theory,
both our approach in this paper and the conventional functional method
for construction of path integrals yield same results as far as for the 
zeroth order of perturbation, i.e. free fields. Since they share same
generating functional and propagators for fundamental fields, they yield
entirely same results for any perturbative calculations. In this sense
they are equivalent. On this point, we should recall that a path
integral is a definite integration and if two definite integrations
share a same answer they are equivalent. Hence the existence of a
changes of variables that brings our formulation to the conventional one
will be expected. Nevertheless, when we need to extract some information
on the state vectors, the advantage of our prescription is evident
because it is built upon manifestly covariant canonical formalism.

In addition to the formulation of path integral, we have found an
interesting operator that yields BRS-inversion and also plays the key
for obtaining an explicit expression of Kugo-Ojima projection in terms
of field variables or creation annihilation operators. The operator
$\hat{Q}_{D}$, that is conserved only in the Feynman gauge, together
with the BRS charge $\hat{Q}_{B}$ provides a nice understanding for
BRS-quartet and their anticommutator is the essential part for the
operator expression of Kugo-Ojima projection. The existence of such an
operator and the solubility of the model will be a consequence of the
topological nature of the quantized system because the genuine physical
state in our model is the vacuum alone. From the view point of explicit
construction, however, it was the viewpoint of quantization according to
the decomposition of the gauge field into physical and unphysical
degrees that made it easy to classify the state vectors in the Fock
space as has been shown in section \ref{sec:kugoojima} because otherwise
we had never met an idea to make use of creation and annihilation
operators, $\hat{\mathcal{B}}$, $\hat{\mathcal{D}}$ and their
conjugates. There will be no other approaches that exhibits in such a
clear way the existence of almost complete analogy between unphysical
degrees of the gauge field and FP ghosts. Hence we may conclude our
prescription for quantization of these variables will be fundamental for
understanding a BRS invariant system. In this regard, non-abelian
generalization of such decomposition will be desired for
non-perturbative analysis of non-abelian gauge theories in terms of
Heisenberg operators. 

Although we have only dealt with the Feynman gauge, considerations on
the discrepancy between the results reported in Ref.~\citen{KS} \ and
Ref.~\citen{AFIO} \ were made to confirm that a path integral constructed
from covariant canonical formalism does not appear to be manifestly
covariant. Hence differs from the conventional manifestly covariant path
integral. It seems that we should adopt the prescription given by
Ref.~\citen{KS} \ to preserve a clear relation to the operator formalism
while having a manifestly covariant expression at the same time.
The significance may not be, however,
always in the appearance of a path integral; The covariance can be
restored from path integrals with non-covariant actions. We must
recognize the importance of the role played by external sources in this
regard as is emphasized by Kashiwa in the first of Ref.\citen{KS}. \  

\section*{Acknowledgements}
The author would like to thank Professor K. Odaka for his continuous
encouragement. He also wishes to express his gratitude to Professor
M. Omote for helpful discussions on several points of this paper. His
special thanks are due to Professor T. Kashiwa for many valuable
comments and suggestions.

\end{document}